\documentclass[preprint,pre,showpacs,showkeys]{revtex4}%
\usepackage{amsfonts}
\usepackage{amsmath}
\usepackage{amssymb}
\usepackage{graphicx}%
\begin{document}
\preprint{Preprint }
\title[The anisotropic XY model on the inhomogeneous periodic chain]{The anisotropic XY model on the inhomogeneous periodic chain}
\author{J. P. de Lima}
\affiliation{Departamento de F\'{\i}sica,Universidade Federal do Piau\'{\i}, Campus
Ministro Petr\^{o}nio Portela,64049-550, Teresina, Piau\'{\i}, Brazil.}
\author{L. L. Gon\c{c}alves}
\affiliation{Departamento de Engenharia Metal\'{u}rgica e de Materiais, Universidade
Federal do Cear\'{a}, Campus do Pici, Bloco 714, 60455-760, Fortaleza,
Cear\'{a}, Brazil}
\email{lindberg@fisica.ufc.br}
\author{T. F. A. Alves}
\affiliation{Departamento de F\'{\i}sica, Universidade Federal do Cear\'{a}, Campus do
Pici, C.P.6030, 60451-970, Fortaleza, Cear\'{a}, Brazil}

\begin{abstract}
The static and dynamic properties of the anisotropic XY-model $(s=1/2)$ on the
inhomogeneous periodic chain, composed of $N$ cells with $n$ different
exchange interactions and magnetic moments, in a transverse field $h,$ are
determined exactly at arbitrary temperatures. The properties are obtained by
introducing the Jordan-Wigner fermionization and by reducing the problem to a
diagonalization of a finite matrix of $nth$ order. The quantum transitions are
determined exactly by analyzing, as a function of the field, the induced
magnetization $1/n\sum_{m=1}^{n}\mu_{m}\left\langle S_{j,m}^{z}\right\rangle $
( $j$ denotes the cell, $m$ the site within the cell, $\mu_{m}$ the magnetic
moment at site $m$ within the cell) and the spontaneous magnetization
$1/n\sum_{m=1}^{n}\left\langle S_{j,m}^{x},\right\rangle $ which is obtained
from the correlations $\left\langle S_{j,m}^{x}S_{j+r,m}^{x}\right\rangle $
for large spin separations. These results, which are obtained for infinite
chains, correspond to an extension of the ones obtained by Tong and
Zhong(\textit{Physica B} \textbf{304, }91 (2001)). The dynamic correlations,
$\left\langle S_{j,m}^{z}(t)S_{j^{\prime},m^{\prime}}^{z}(0)\right\rangle $,
and the dynamic susceptibility, $\chi_{q}^{zz}(\omega),$ are also obtained at
arbitrary temperatures. Explicit results are presented in the limit $T=0$,
where the critical behaviour occurs, for the static susceptibility $\chi
_{q}^{zz}(0)$ as a function of the transverse field $h$, and for the frequency
dependency of dynamic susceptibility $\chi_{q}^{zz}(\omega)$.

\end{abstract}
\date{\today}

\pacs{05.70.Fh%
$\backslash$
05.70.Jk%
$\backslash$
75.10.Jm%
$\backslash$
75.10.Pq}
\keywords{anisotropic XY-model, quantum transition, inhomogeneous chain}\eid{identifier}
\maketitle

\section{Introduction}

The one-dimensional XY-model introduced by Lieb, Schultz and Mattis
\cite{lieb:1961} is still one of the few quantum many-body problems \ that can
be solved exactly. Although almost forty-five years have passed since the
original solution has been proposed, this old model continues to provide new
information on the quantum behaviour of magnetic
systems\cite{dagotto:1996,nguyen:1996,gambardella:2002,mukherjeea:2004,matsumoto:2004}%
. In particular, it has shed some light on the quantum phase transitions
\cite{sachdev:2000}, since exact results can be obtained for most of its
properties. The model has also been applied in the study of quantum
entanglement, which plays an essential role in the quantum computation.
Important results of this application can be found in the recent work by Amico
et al.\cite{Amico:2006} \ and in the references therein.

In this paper we will consider the anisotropic XY-model in a transverse field
on the inhomogeneous periodic (closed) chain consisting of $N$ cells composed
\ of $n$ sites, whose version on the open chain has been recently addressed by
Feldman\cite{feldman:2006}. It has also been studied by Tong and Zhong
\cite{tongzhong:2001} and Derzhko et al.\cite{derzhko:2004}, who have
restricted their analysis to the study of the thermodynamic properties. To the
best of our knowledege, the work \ by Tong and Liu\cite{tongliu:2006}, which
concerns the study of the zeros of the partition function, at T=0, and its
relation to the quantum phase transitions contains the latest results on the model.

The anisotropic model corresponds to an extension of the isotropic one
recently studied by the authors\cite{delima:2006}. Its study on the
inhomogeneous periodic chain also corresponds to an extension of its version
on the alternating
superlattice\cite{siskens:1975,perk:1975,perk:1980,taylor:1985,barbosafilho:2001,delima:2002}%
, and, in this work, we will solve exactly the model by considering $n$
different exchange constants and magnetic moments. The aim of this work is to
present a comprehensive study of its static and dynamic quantum critical behaviour.

In Section 2 we introduce the model and diagonalize its Hamiltonian. An
explicit expression is presented for $n=2,$ and the results compared to the
known ones obtained by various authors
\cite{siskens:1975,perk:1975,perk:1980,taylor:1985,barbosafilho:2001,delima:2002}%
. We also present the solution of the model for $n=8$, which has been obtained numerically.

The induced magnetization $M^{z}$ and \ the isothermal susceptibility
$\chi_{T}^{zz},$ at arbitrary temperature, are obtained in Section 3, and
explicit expressions presented for $T=0$. By analyzing these quantities, the
quantum critical behaviour is studied and the critical exponents associated to
the quantum transitions are obtained.

The spontaneous magnetization $M^{x}$ is obtained in Section 4, from the
two-spin correlation $\left\langle S_{l,m}^{x}S_{l+r,m}^{x}\right\rangle ,$
and the critical behaviour is also obtained, with high accuracy, numerically,
by means of a non-linear regression of the data evaluated for finite values of
$r$. This rather surprising result allows for the complete determination of
$M^{x}$ as a function of the field, and to determine the multiple quantum
transitions undergone by the model.

The static and dynamic correlations $\left\langle \tau_{l}^{z}\tau_{l+r}%
^{z}\right\rangle $ are presented in Section 5 and, the dynamic susceptibility
$\chi_{q}^{zz}(\omega),$ in Section 6. Finally, in Section 7, we summarize the
main results of the paper.

\section{The model}

We consider the anisotropic XY- model $(s=1/2)$ on the inhomogeneous periodic
chain with $N$ cells, $n$ sites per cell, and lattice parameter $a$, in a
transverse field, whose unit cell is shown in Fig. 1, which corresponds to an
extension of the isotropic model recentely considered \cite{delima:2006}. The
Hamiltonian is given by
\begin{align}
H  &  =-\sum_{l=1}^{N}\left\{  \sum_{m=1}^{n}\mu_{m}hS_{l,m}^{z}+\left[
\sum_{m=1}^{n-1}J_{m}\left(  1+\gamma_{m}\right)  S_{l,m}^{x}S_{l,m+1}%
^{x}+J_{m}\left(  1-\gamma_{m}\right)  S_{l,m}^{y}S_{l,m+1}^{y}\right]
\right.+ \nonumber\\
&  \left.  +J_{n}\left(  1+\gamma_{n}\right)  S_{l,n}^{x}S_{l+1,1}^{x}%
+J_{n}\left(  1-\gamma_{n}\right)  S_{l,n}^{y}S_{l+1,1}^{y}\right\}  ,
\label{HamiltonianaXY}%
\end{align}

%

\begin{figure}
[ptb]
\begin{center}
\includegraphics[width=0.5\columnwidth
]%
{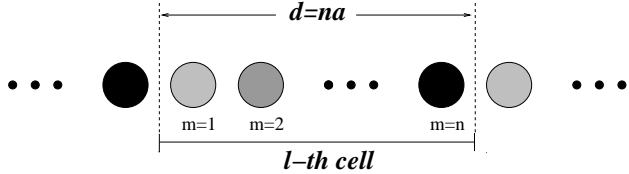}%
\caption{Unit cell of the inhomogeneous periodic chain}%
\label{fig1}%
\end{center}
\end{figure}

\noindent where the parameters $J_{l,m}$ are the exchange couplings between
nearest-neighbour, $\mu_{m}$ the magnetic moments, $h$ the external field, and
we have assumed periodic boundary conditions. If we introduce the ladder
operators%
\begin{equation}
S_{l,m}^{\pm}=S_{l,m}^{x}\pm iS_{l,m}^{y}, \label{ladder}%
\end{equation}
and the generalized Jordan-Wigner transformation\cite{goncalves:1995}
\begin{align}
S_{l,m}^{^{+}}  &  =\exp\left\{  i\pi\sum_{l^{\prime}=1}^{l-1}\sum_{m^{\prime
}=1}^{n}c_{l^{\prime},m^{\prime}}^{\dagger}c_{l^{\prime},m^{\prime}}+i\pi
\sum_{m^{\prime}=1}^{m-1}c_{l,m^{\prime}}^{\dagger}c_{l,m^{\prime}}\right\}
c_{l,m}^{\dagger},\nonumber\\
S_{l,m}^{z}  &  =c_{l,m}^{\dagger}c_{l,m}-\frac{1}{2}, \label{Jordan-Wigner}%
\end{align}
where $c_{l,m}$ and $c_{l,m}^{\dagger}$ are fermion annihilation and creation
operators, we can write the Hamiltonian as\cite{siskens:1974}
\begin{equation}
H=H^{+}P^{+}+H^{-}P^{-}, \label{HamiltonianFermions}%
\end{equation}
where%
\begin{align}
H^{\pm}  &  =-\sum_{l=1}^{N}\left\{  \sum_{m=1}^{n}\mu_{m}h\left(
c_{l,m}^{\dagger}c_{l,m}-\frac{1}{2}\right)  \right.  +\sum_{m=1}^{n-1}\left[
\frac{J_{m}}{2}\left(  c_{l,m}^{\dagger}c_{l,m+1}+c_{l,m+1}^{\dag}%
c_{l,m}\right)  \right.  +\nonumber\\
&  \left.  \left.  +\frac{J_{m}\gamma_{m}}{2}\left(  c_{l,m}^{\dagger
}c_{l,m+1}^{\dagger}+c_{l,m+1}c_{l,m}\right)  \right]  \right\}  -\sum
_{l=1}^{N-1}\left[  \frac{J_{n}}{2}\left(  c_{l,n}^{\dagger}c_{l+1,1}%
+c_{l+1,1}^{\dagger}c_{l,n}\right)  \right.  +\nonumber\\
&  \left.  +\frac{J_{n}\gamma_{n}}{2}\left(  c_{l,n}^{\dagger}c_{l+1,1}%
^{\dagger}+c_{l+1,1}c_{l,n}\right)  \right]  \pm\left[  \frac{J_{n}}{2}\left(
c_{N,n}^{\dagger}c_{1,1}+c_{1,1}^{\dagger}c_{N,n}\right)  +\frac{J_{n}%
\gamma_{n}}{2}\left(  c_{N,n}^{\dagger}c_{1,1}^{\dagger}+c_{1,1}%
c_{N,n}\right)  \right]  ,
\end{align}
and%
\begin{equation}
P^{\pm}=\frac{I\pm P}{2},
\end{equation}
with $P$ given by
\begin{equation}
P=\exp\left(  i\pi\sum_{l=1}^{N}\sum_{m=1}^{n}c_{l,m}^{\dagger}c_{l,m}\right)
.
\end{equation}

As it is well known \cite{siskens:1974,capel:1977,goncalvestese:1977}, since
the operator $P$ commutes with the Hamiltonian, the eigenstates have definite
parity, and $P^{-}(P^{+})$ corresponds to a projector into a state of odd
(even) parity.

Introducing periodic and anti-periodic boundary conditions on $c^{\prime}s$
for $H^{-}$ and $H^{+}$ respectively, the wave-vectors in the Fourier
transform \cite{barbosafilho:2001},
\begin{align}
c_{l,m}  &  =\frac{1}{\sqrt{N}}\sum_{q}\exp\left(  -iqdl\right)
a_{q,m},\nonumber\\
a_{q,m}  &  =\frac{1}{\sqrt{N}}\sum_{l=1}^{N}\exp\left(  iqdl\right)  c_{l,m},
\label{fourierXY}%
\end{align}
are given by $q^{-}=\frac{2l\pi}{Nd},$ for periodic condition and $q^{+}%
=\frac{\pi(2l+1)}{Nd}$, for anti-periodic condition, with $l=0,\pm1,....,\pm
N/2$, and $H^{-}$ and $H^{+}$ can be written in the form%
\begin{equation}
H^{\pm}=\sum_{q^{\pm}}H_{q^{\pm}}, \label{Hamiltonianq}%
\end{equation}
where
\begin{align}
H_{q^{\pm}}  &  =-\sum_{m=1}^{n}\mu_{m}h\left(  a_{q^{\pm},m}^{\dagger
}a_{q^{\pm},m}-\frac{1}{2}\right)  -\nonumber\\
&  -\sum_{m=1}^{n-1}\frac{J_{m}}{2}\left[  a_{q^{\pm},m}^{\dagger}a_{q^{\pm
},m+1}+a_{q^{\pm},m+1}^{\dagger}a_{q^{\pm},m}+\gamma_{m}\left(  a_{q^{\pm}%
,m}^{\dagger}a_{q^{\pm},m+1}^{\dagger}+a_{q^{\pm},m+1}a_{q^{\pm},m}\right)
\right]  -\nonumber\\
&  -\frac{J_{n}}{2}\left[  a_{q^{\pm},n}^{\dagger}a_{q^{\pm},1}\exp(-idq^{\pm
})+a_{q^{\pm},1}^{\dagger}a_{q^{\pm},n}\exp(idq^{\pm})\right.  +\nonumber\\
&  +\left.  \gamma_{m}\left(  a_{q^{\pm},n}^{\dagger}a_{q^{\pm},1}^{\dagger
}\exp(-idq^{\pm})+a_{q^{\pm},1}a_{q^{\pm},n}\exp(idq^{\pm})\right)  \right]  .
\label{Hq}%
\end{align}
Although $H^{-}$ and $H^{+}$ do not commute, it can be shown that in the
thermodynamic limit all the static properties of the \ system can be obtained
in terms of $H^{-}$ or $H^{+}.$ However, even in this limit, some dynamic
properties depend on $H^{-}$ and $H^{+}$
\cite{siskens:1974,capel:1977,goncalvestese:1977}. Since $\left[  H_{q^{\pm}%
},H_{-q^{\pm}}\right]  \neq0,$ we consider the symmetrization%
\begin{equation}
\widetilde{H}_{q^{\pm}}\equiv H_{q^{\pm}}+H_{-q^{\pm}}, \label{symmetrization}%
\end{equation}
and the Hamiltonian $H^{\pm}$ can be written as
\begin{equation}
H^{\pm}=\frac{1}{2}%
{\displaystyle\sum\limits_{q^{\pm}}}
\widetilde{H}_{q^{\pm}}.
\end{equation}
By making the identification $q^{\pm}\equiv q,$ it follows immediately that
$\left[  \widetilde{H}_{q},\widetilde{H}_{q^{\prime}}\right]  =0.$ Therefore,
following the procedure introduced by Lieb, Schultz and Mattis \cite{lieb:1961}%
, we can diagonalize the Hamiltonian $H^{\pm}$ by introducing the canonical
transformation%
\begin{align}
a_{q,m} &  =\sum\limits_{k}^{n}\left(  \frac{\psi_{q,k,m}^{\ast}+\phi
_{q,k,m}^{\ast}}{2}\eta_{q,k}+\frac{\psi_{q,k,m}^{\ast}-\phi%
_{q,k,m}^{\ast}}{2}\eta_{-q,k}^{\dagger}\right)  ,\nonumber\\
a_{q,m}^{\dagger} &  =\sum\limits_{k}^{n}\left(  \frac{\psi_{q,k,m}%
+\phi_{q,k,m}}{2}\eta_{q,k}^{\dagger}+\frac{\psi_{q,k,m}%
-\phi_{q,k,m}}{2}\eta_{-q,k}\right) \label{tcanonical} ,
\end{align}
where $\mathbf{\phi}_{q,k,m}$ and $\mathbf{\psi}_{q,k,m}$ are the components of the  eigenvectors, $\mathbf{\Phi}_{q,k}$ and $\mathbf{\Psi}_{q,k}$,
of the matrices $\left(  \mathbb{A}_{q}-\mathbb{B}_{q}\right)  \left(
\mathbb{A}_{q}+\mathbb{B}_{q}\right)  $ and $\left(  \mathbb{A}_{q}%
+\mathbb{B}_{q}\right)  \left(  \mathbb{A}_{q}-\mathbb{B}_{q}\right)  ,$ with
$\mathbb{A}_{q}$ and $\mathbb{B}_{q}$ given by%

\begin{equation}
\mathbb{A}_{q}=-\left(
\begin{array}
[c]{llllll}%
\mu_{1}h & \frac{J_{1}}{2} & 0 & \cdots & 0 & \frac{J_{n}}{2}\exp\left(
iqd\right)  \\
\frac{J_{1}}{2} & \mu_{2}h & \frac{J_{2}}{2} &  &  & 0\\
0 & \frac{J_{2}}{2} & \mu_{3}h & \frac{J_{3}}{2} &  & \vdots\\
\vdots &  & \frac{J_{3}}{2} & \ddots & \ddots & 0\\
0 &  &  & \ddots & \mu_{n-1}h & \frac{J_{n-1}}{2}\\
\frac{J_{n}}{2}\exp\left(  -iqd\right)   & 0 & \cdots & 0 & \frac{J_{n-1}}{2}
& \mu_{n}h
\end{array}
\right)  ,\label{Aq}%
\end{equation}%
\begin{equation}
\mathbb{B}_{q}=\left(
\begin{array}
[c]{llllll}%
0 & -\frac{J_{1}\gamma_{1}}{2} & 0 & \cdots & 0 & \frac{J_{n}\gamma_{n}}%
{2}\exp\left(  iqd\right)  \\
\frac{J_{1}\gamma_{1}}{2} & 0 & -\frac{J_{2}\gamma_{2}}{2} &  &  & 0\\
0 & \frac{J_{2}\gamma_{2}}{2} & 0 & -\frac{J_{3}\gamma_{3}}{2} &  & \vdots\\
\vdots &  & \frac{J_{3}\gamma_{3}}{2} & \ddots & \ddots & 0\\
0 &  &  & \ddots & 0 & -\frac{J_{n-1}\gamma_{n-1}}{2}\\
-\frac{J_{n}\gamma_{n}}{2}\exp\left(  -iqd\right)   & 0 & \cdots & 0 &
\frac{J_{n-1}\gamma_{n-1}}{2} & 0
\end{array}
\right)  .\label{Bq}%
\end{equation}
The corresponding eigenvalues are $E_{q,k}^{2},$ the squares of the fermion energy
levels, such that the Hamiltonian $H^{\pm},$ given in eq.(\ref{Hq}), can be
written in the diagonal form%
\begin{equation}
H^{\pm}=\sum_{q,k}E_{q,k}\left(  \eta_{q,k}^{\dagger}\eta_{q,k}-\frac{1}%
{2}\right)  , \label{diagonalform}%
\end{equation}
where for $H^{+}(H^{-})$ the wave-vector $q$ is identical to $q^{+}(q^{-}),$
and the spectrum $E_{q,k}$ of $\ \widetilde{H}_{q}$ is determined from the
determinantal equation
\begin{equation}
\det\left[  \left(  \mathbb{A}_{q}+\mathbb{B}_{q}\right)  \left(
\mathbb{A}_{q}-\mathbb{B}_{q}\right)  -E_{q}^{2}\mathbb{I}\right]  =0.
\label{spectrum}%
\end{equation}

Since we are interested in the thermodynamic limit, we will consider from here
on $H^{-}$ only, in all calculations.

Explicit expressions for the excitation spectrum can be determined for $n$
equal to 2, 3 and 4, and for $n$ greater than 4, it is determined numerically
from eq.(\ref{spectrum}). Since we are considering the particle-hole
representation, this implies that $E_{q}\geqslant0.$

Explicitly, for $n=2$, the excitation spectrum is given by%

\begin{align}
E_{q}^{2}  &  =\frac{\alpha_{1}+\alpha_{2}+(\lambda_{1}+\lambda_{2})\cos2q}%
{2}\pm\nonumber\\
&  \frac{1}{2}\sqrt{\left[  \alpha_{1}-\alpha_{2}+(\lambda_{1}-\lambda
_{2})\cos2q\right]  ^{2}+4\left[  \delta_{1}^{2}+\delta_{2}^{2}+2\delta
_{1}\delta_{2}\cos(2q)\right]  }, \label{spectrum2}%
\end{align}

where%

\begin{align}
\alpha_{1}  &  =(\mu_{1}h)^{2}+\frac{J_{1}^{2}}{4}(1+\gamma_{1})^{2}%
+\frac{J_{2}^{2}}{4}(1-\gamma_{2})^{2},\nonumber\\
\alpha_{2}  &  =(\mu_{2}h)^{2}+\frac{J_{2}^{2}}{4}(1+\gamma_{2})^{2}%
+\frac{J_{1}^{2}}{4}(1-\gamma_{1})^{2},\nonumber\\
\lambda_{1}  &  =\frac{J_{1}J_{2}}{2}(1+\gamma_{1})(1-\gamma_{2}),\nonumber\\
\lambda_{2}  &  =\frac{J_{1}J_{2}}{2}(1-\gamma_{1})(1+\gamma_{2}),\nonumber\\
\delta_{1}  &  =\frac{J_{1}}{2}(1-\gamma_{1})\mu_{1}h+\frac{J_{1}}{2}%
(1+\gamma_{1})\mu_{2}h,\nonumber\\
\delta_{2}  &  =\frac{J_{2}}{2}(1+\gamma_{2})\mu_{1}h+\frac{J_{2}}{2}%
(1-\gamma_{2})\mu_{2}h.
\end{align}

For $\gamma_{1}=\gamma_{2}=0,$ the previous result reproduces the excitation
spectrum obtained for the isotropic case \cite{delima:2006}, provided we
consider the particle representation which allows for negative energies. It
also reproduces the result obtained by Siskens et al.\cite{siskens:1975}.

As already pointed out, although explicit expressions can be determined for
$n$ equal to 3 and 4, they are not presented here since they are too cumbersome.

Even in the case where we have identical magnetic moments, differently from
the isotropic case, the effect of the the field does not correspond to a
translation of the excitation spectrum for zero field, since the term of the
field does not commute with the Hamiltonian.

The critical fields which characterize the quantum transitions are determined
from the excitation spectrum by imposing the condition $E_{q}=0$ for $q=0$ and
$q=\pi/d$\cite{delima:2006}. The number of transitions is highly dependent on
the anisotropy and varies from $n$ to one. In particular, the limit of a
single transition always occurs for the transverse Ising model which
corresponds to the $\gamma^{\prime}s$ equal to one.

\begin{figure}
[ptb]
\begin{center}
\includegraphics[width=0.5\columnwidth
]%
{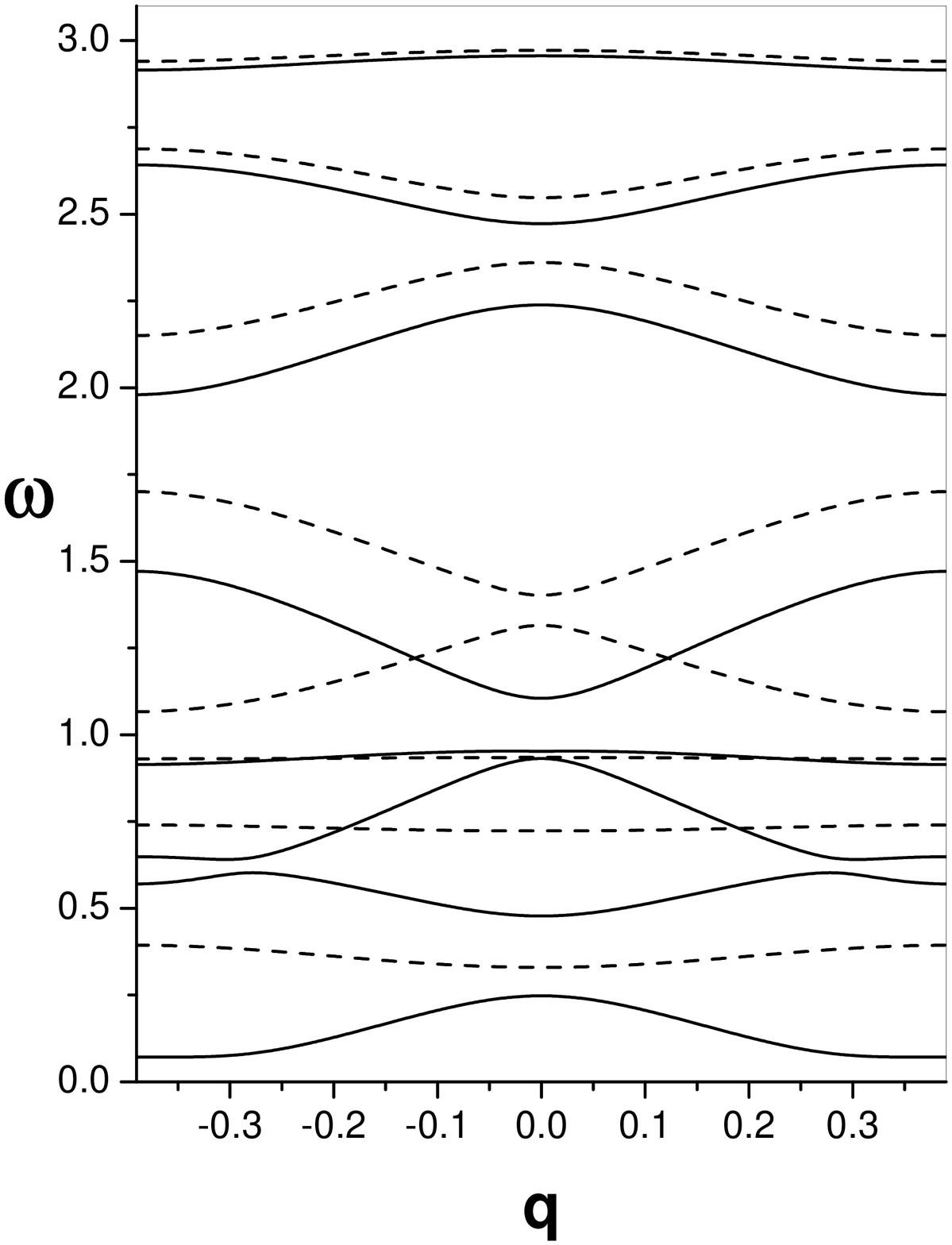}%
\caption{Excitation spectrum for $n=8,J_{1}=J_{6}=J_{8}=1.5,$ $J_{2}=3,$
$J_{3}=1.5,$ $J_{4}=$ $J_{5}=J_{7}=2,$ $h=1$, $\mu_{1}=...=\mu_{8}=1$ and
$\gamma_{1}=...=\gamma_{8}=0.1$ (continuous line), and for $\gamma_{1}=0.1,$
$\gamma_{2}=0.2,$ $\gamma_{3}=0.3,$ $\gamma_{4}=0.4,\gamma_{5}=0.5,$
$\gamma_{6}=0.6,\gamma_{7}=0.7,$ $\gamma_{8}=0.8$ (dashed line).}%
\label{fig2}%
\end{center}
\end{figure}

The excitation spectrum for $n=8$ and identical magnetic moments is presented
in Fig.\ref{fig2} for identical and different $\gamma^{\prime}s$, where we
have considered the lattice spacing $a=1.$ As it can be seen, the energy of
the modes is always positive and the effect of the spatial variation of the
anisotropy is more pronounced in the low energy modes. Since there is no zero
energy excitation, the model is not critical at the considered value of the
transverse field.%

\section{The induced magnetization and isothermal susceptibility $\chi
_{T}^{zz}$}

From eqs.(\ref{Jordan-Wigner}), the operator $S_{l,m}^{z}$ can be written as%
\begin{equation}
S_{l,m}^{z}=-\frac{1}{2}A_{l,m}B_{l,m}, \label{Szlm}%
\end{equation}
with
\begin{align}
A_{l,m}  &  \equiv c_{l,m}^{\dagger}+c_{l,m},\nonumber\\
B_{l,m}  &  \equiv c_{l,m}^{\dagger}-c_{l,m}, \label{comblinferm}%
\end{align}
which can be expressed, from eqs.(\ref{fourierXY})and (\ref{tcanonical}), as
\begin{align}
A_{l,m}  &  =\frac{1}{\sqrt{N}}\sum_{q,k}e^{iqdl}\psi_{q,k,m}\left(
\eta_{q,k}^{\dagger}+\eta_{-q,k}\right)  ,\nonumber\\
B_{l,m}  &  =\frac{1}{\sqrt{N}}\sum_{q,k}e^{iqdl}\phi_{q,k,m}\left(
\eta_{q,k}^{\dagger}-\eta_{-q,k}\right)  . \label{AeB}%
\end{align}
By using eqs.(\ref{Szlm}) and (\ref{AeB}), the local magnetization
$M_{l,m}^{z}$ can be written as%
\begin{equation}
M_{l,m}^{z}\equiv\mu_{m}\left\langle S_{l,m}^{z}\right\rangle =-\frac{1}%
{2N}\sum\limits_{q,k}\mu_{m}\psi_{q,k,m}\phi_{q,k,m}^{\ast}\left(
1-2n_{q,k}\right)  , \label{localmagnetization}%
\end{equation}
where the fermion occupation number $n_{q,k}$ is given by%
\begin{equation}
n_{q,k}=\frac{1}{1+e^{\overline{E}_{q,k}}}, \label{occupationnumber}%
\end{equation}
where $\overline{E}_{q,k}=\beta E_{q,k}\equiv E_{q,k}/k_{B}T$, and the
calculation has been done by considering $H=H^{-}$
\cite{siskens:1974,capel:1977,goncalvestese:1977}, since we are interested in
the thermodynamic limit.

As in the isotropic model \cite{delima:2006}, we define an average cell
magnetization operator in the $z$ direction, $\tau_{l}^{z},$ as%
\begin{equation}
\tau_{l}^{z}\equiv\frac{1}{n}\sum_{m=1}^{n}\mu_{m}S_{l,m}^{z},
\label{operatortauzlm}%
\end{equation}
then the induced magnetization per site, $M^{z},$ is equal to $\left\langle
\tau_{l}^{z}\right\rangle $ and, from eq.(\ref{localmagnetization}), can be
written in the form%
\begin{equation}
M^{z}\equiv\left\langle \tau_{l}^{z}\right\rangle =-\frac{1}{2nN}%
\sum\limits_{q,k,m}\mu_{m}\psi_{q,k,m}\phi_{q,k,m}^{\ast}\tanh\left(
\frac{^{\overline{E}_{q,k}}}{2}\right)  , \label{magnetization}%
\end{equation}
which at $T=0$ becomes
\begin{equation}
M^{z}=-\frac{1}{2nN}\sum\limits_{q,k,m}\mu_{m}\psi_{q,k,m}\phi_{q,k,m}^{\ast}.
\label{magnetizationT=0}%
\end{equation}
The isothermal susceptibility can be obtained from eqs.(\ref{magnetization})
and (\ref{magnetizationT=0}) by means of the expression
\begin{equation}
\chi_{T}^{zz}\equiv\frac{1}{n}\frac{\partial M^{z}}{\partial h},
\label{Isothermal Suceptibility Chizz}%
\end{equation}

\noindent which has to be evaluated numerically.%

\begin{figure}
[ptb]
\begin{center}
\includegraphics[width=0.5\columnwidth
]%
{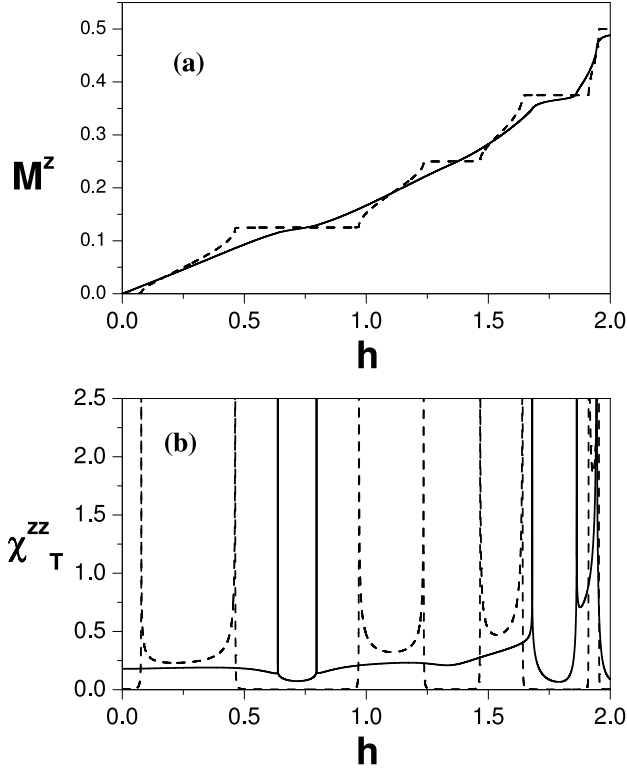}%
\caption{(a) Magnetization $M^{z}$ and (b) isothermal susceptibility $\chi
_{T}^{zz},$at $T=0,$ as functions of the field for $n=8,$ $\mu_{1}=...=\mu
_{8}=1$, $J_{1}=J_{6}=J_{8}=1.5,$ $J_{2}=3,$ $J_{3}=1.5,$ $J_{4}=$
$J_{5}=J_{7}=2,$ $\gamma=0.1$(dashed line) and for $\gamma=0.15$(continuous
line).}%
\label{fig3}%
\end{center}
\end{figure}

\begin{figure}
[ptb]
\begin{center}
\includegraphics[width=0.5\columnwidth
]%
{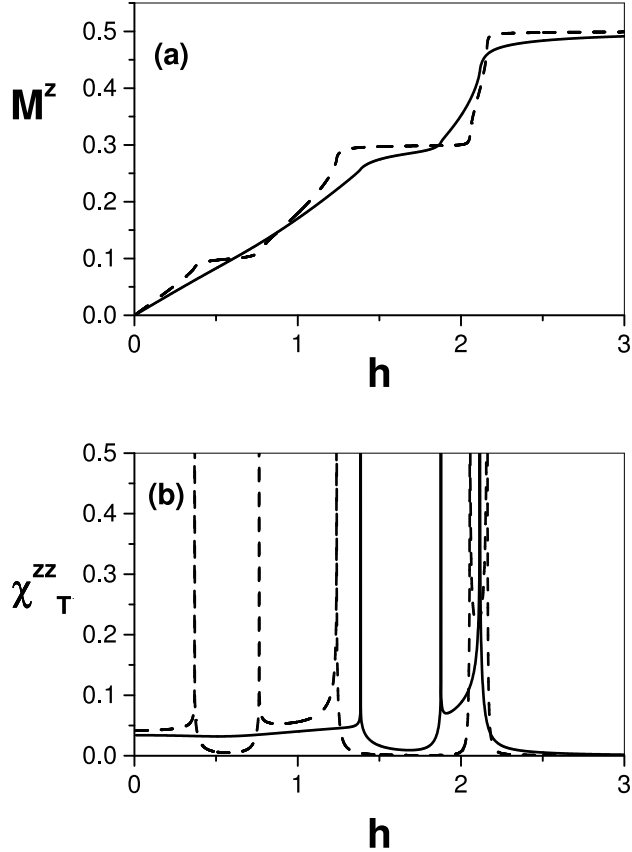}%
\caption{(a) Magnetization $M^{z}$ and (b) isothermal susceptibility $\chi
_{T}^{zz},$at $T=0,$ as functions of the field for $n=5,$ $\mu_{1}=...=\mu
_{5}=1$, $J_{1}=1,J_{2}=1.5,$ $J_{3}=2,$ $J_{4}=2.5$ $J_{5}=3,$ $\gamma
=0.1$(dashed line) and for $\gamma=0.3$(continuous line).}%
\label{fig4}%
\end{center}
\end{figure}

At $T=0,$ the isothermal susceptibility $\chi_{T}^{zz}$ diverges at the
quantum critical points induced by the field. This result can be seen in
Fig.\ref{fig3} where $M^{z}$ and $\chi_{T}^{zz}$ are presented as functions of
the field $h,$ at $T=0$, for a chain with $n=8$, identical $\mu$'s,
$\gamma=0.1$ and $0.15$. The magnetization, differently from the isotropic
model, does not present plateaus but it does present inflexion points which
induce the divergences of the susceptibility at the quantum phase transitions.
As expected, the results also show that for $n$ even there is the tendency of
formation of a zero magnetization plateau \ in the limit $\gamma\rightarrow0$,
which is not present for $n$ odd as shown in Fig.\ref{fig4} for $n=5$.%

\begin{figure}
[ptb]
\begin{center}
\includegraphics[width=0.5\columnwidth
]%
{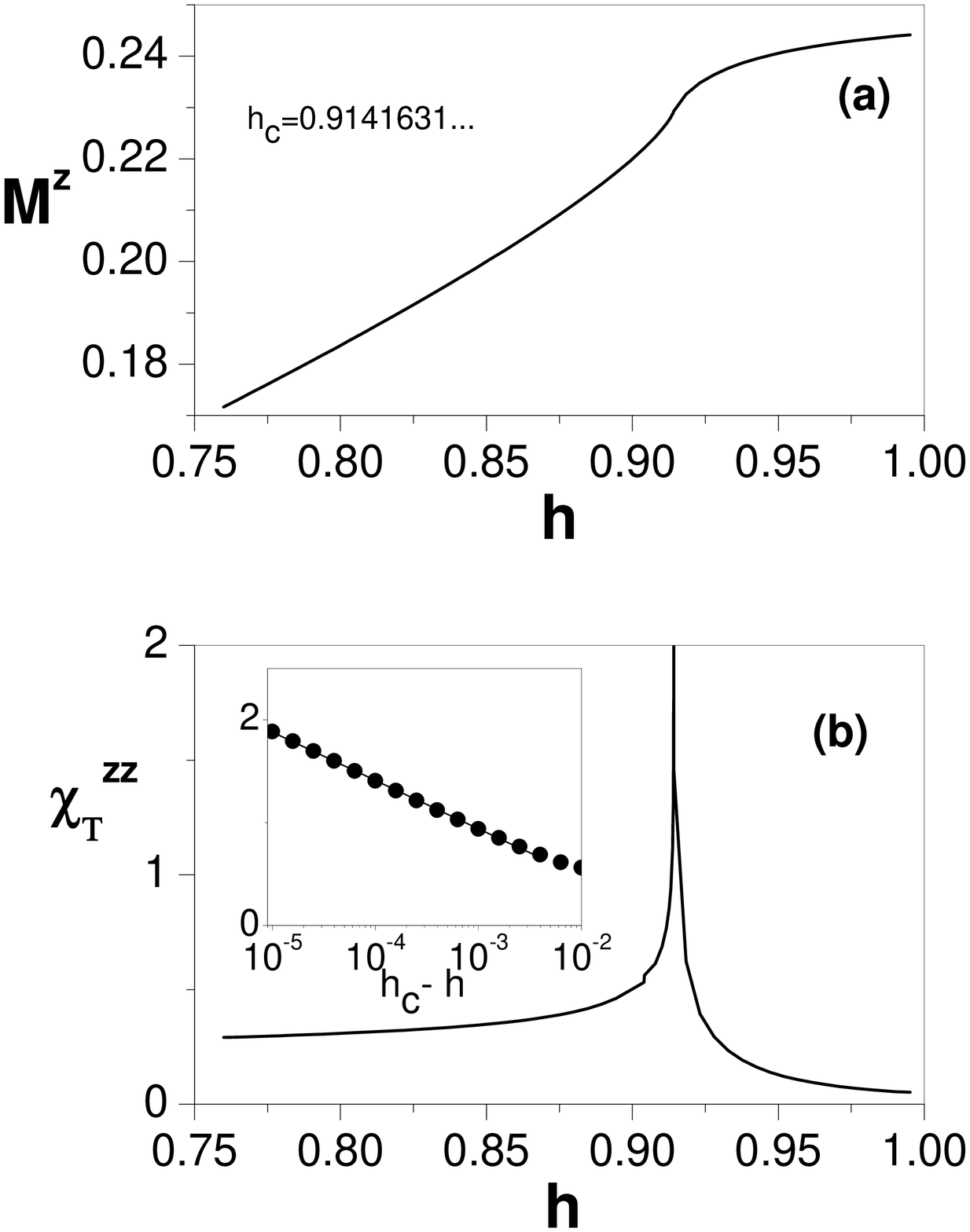}%
\caption{(a) Magnetization M$^{z}$ and (b) isothermal susceptibility $\chi
_{T}^{zz},$ at $T=0$ and $n=4,$ $\mu_{1}=\mu_{2}=\mu_{3}=\mu_{4}=1$,
$\gamma_{1}=\gamma_{2}=\gamma_{3}=\gamma_{4}=0.1,$ as functions of the uniform
field for $J_{1}=1,$ $J_{2}=1.5,$ $J_{3}=1.75,$ $J_{4}=2.0.$ The logarithmic
singular behaviour of $\chi_{T}^{zz}$ it is shown in the inset in (b) where the horizontal scale is decimal logarithmic. }%
\label{fig5}%
\end{center}
\end{figure}

At $T=0,$ on the critical region, we have verified numerically that the
isothermal susceptibility presents the behavior%
\begin{equation}
\chi_{T}^{zz}\sim\ln\left\vert h-h_{c}\right\vert \quad(0<\gamma_{m}\leq1),
\end{equation}
as it can be seen in the results presented in Fig.\ref{fig5}, for a chain with
$n=4$, and identical $\mu^{\prime}$s and $\gamma^{\prime}s.$%

\bigskip

The transitions belong to the same universality class as the one in the
homogeneous model \cite{niemeijer:1967}. For $T>0$ all these transitions are
suppressed by the thermal fluctuations.

As already pointed out, the critical fields are associated with the
zero-energy modes with $q=0$ and $q=\pi/d,$ and the number of critical points
can vary from one in the transverse Ising model limit, for equal
$\gamma^{\prime}s=1,$ to $n,$ provided the $\mu^{\prime}s$ are positive. On
the other hand, if we allow the $\mu^{\prime}s$ to assume negative values, we
can have more than one phase transition even in the transverse Ising model
limit, as has been shown by Derzhko et al.\cite{derzhko:2004}.

Then, the critical fields can be obtained numerically from the excitation
spectrum and the results are shown in Fig.(\ref{fig6}a) for a chain with
$n=4$, and identical $\mu^{\prime}$s and $\gamma^{\prime}s.$ As can be
verified, the number of transitions depends on $\gamma$ and its signature is
present in the behaviour of the spontaneous and induced magnetizations. From
these results, we can also conclude that for $n$ even we have two critical
$\gamma^{\prime}s$, which can induce quantum transitions at zero field,
whereas for $n$ odd we have just a single critical $\gamma$, equal to zero, as
in the uniform model \cite{barouch:1971}. This result is shown in
Fig.(\ref{fig7}) and, differently from the isotropic model, it constitutes the
main effect of the cell size on the quantum critical behaviour of the model.

\section{The magnetization $M^{x}$}

Since the Hamiltonian is invariant under the transformation $S_{l,m}%
^{x}\rightarrow-S_{l,m}^{x},$ which means that the $<S_{l,m}^{x}>$ is equal to
zero, we cannot calculate the spontaneous magnetization directly by using the
Hamiltonian shown in eq.(\ref{HamiltonianaXY}). On the other hand, the
Hamiltonian after the introduction of an external field along $x$ direction,
which would eliminate the mentioned symmetry property, is no more exactly
soluble. Therefore, by using the original Hamiltonian, the spontaneous local
magnetization, $M_{m}^{x},$ can be determined, from the static correlation
function $\ \left\langle S_{l,m}^{x}S_{l+r,m}^{x}\right\rangle ,$ by means of
the well known decomposition\cite{stanley:1971}%

\begin{equation}
\lim_{r\rightarrow\infty}\ \left\langle S_{l,m}^{x}S_{l+r,m}^{x}\right\rangle
=\left\langle S_{l,m}^{x}\rangle\langle S_{l+r,m}^{x}\right\rangle .
\label{decomposition}%
\end{equation}

Then we can write immediately the local spontaneous magnetization as%

\begin{equation}
M_{m}^{x}=\lim_{r\rightarrow\infty}\sqrt{\left\langle S_{l,m}^{x}S_{l+r,m}%
^{x}\right\rangle }, \label{corrsx}%
\end{equation}

and from this we obtain the average spontaneous magnetization per cell,
$M^{x},$ given by%

\begin{equation}
M^{x}=\frac{1}{n}\sum_{m=1}^{n}M_{m}^{x}. \label{corrsx-med}%
\end{equation}

Therefore, by using the eqs.(\ref{ladder}), (\ref{Jordan-Wigner}) and
(\ref{comblinferm}), the correlation $\left\langle S_{l,m}^{x}S_{l+r,m}%
^{x}\right\rangle $ can be written as%
\begin{equation}
<S_{l,m}^{x}S_{l+r,m}^{x}>=\frac{1}{4}<B_{l,m}A_{l,m+1}B_{l,m+1}%
A_{l,m+2}B_{l,m+2}\ldots A_{l+r,m-1}B_{l+r,m-1}A_{l+r,m}>.
\label{sxsxcomblinferm}%
\end{equation}
Following \cite{lieb:1961} we can write $\left\langle S_{l,m}^{x}S_{l+r,m}%
^{x}\right\rangle ,$ by using using the Wick's theorem, in terms of the determinant%

\begin{equation}
\left\langle S_{l,m}^{x}S_{l+r,m}^{x}\right\rangle =\frac{1}{4}det\left(
\begin{array}
[c]{cccccc}%
\left\langle B_{l,m}A_{l,m+1}\right\rangle  & \left\langle B_{l,m}%
A_{l,m+2}\right\rangle  & \ldots & \left\langle B_{l,m}A_{l,n}\right\rangle  &
\ldots & \left\langle B_{l,m}A_{l+r,m}\right\rangle \\
\left\langle B_{l,m+1}A_{l,m+1}\right\rangle  & \left\langle B_{l,m+1}%
A_{l,m+2}\right\rangle  & \ldots & \left\langle B_{l,m+1}A_{l,n}\right\rangle
& \ldots & \left\langle B_{l,m+1}A_{l+r,m}\right\rangle \\
\left\langle B_{l,m+2}A_{l,m+1}\right\rangle  & \left\langle B_{l,m+2}%
A_{l,m+2}\right\rangle  & \ldots & \left\langle B_{l,m+2}A_{l,n}\right\rangle
& \ldots & \left\langle B_{l,m+2}A_{l+r,m}\right\rangle \\
\vdots & \vdots & \ddots &  & \ddots & \vdots\\
\left\langle B_{l,n}A_{l,m+1}\right\rangle  & \left\langle B_{l,n}%
A_{l,m+2}\right\rangle  & \ldots & \left\langle B_{l,n}A_{l,n}\right\rangle  &
\ldots & \left\langle B_{l,n}A_{l+r,m}\right\rangle \\
\vdots & \vdots & \ldots & \vdots & \ldots & \vdots\\
\left\langle B_{l+r,m-1}A_{l,m+1}\right\rangle  & \left\langle B_{l+r,m-1}%
A_{l,m+2}\right\rangle  & \ldots & \left\langle B_{l+r,m-1}A_{l+r,n}%
\right\rangle  & \ldots & \left\langle B_{l+r,n}A_{l+r,n}\right\rangle
\end{array}
\right)  , \label{determinante-sx}%
\end{equation}
where the static contractions $\left\langle B_{l,m}A_{l^{\prime},m^{\prime}%
}\right\rangle ,$ at arbitrary $T,$ are given by%

\begin{equation}
\left\langle B_{l,m}A_{l^{\prime},m^{\prime}}\right\rangle =\frac{1}{N}%
\sum_{q,k}e^{-iqd(l^{\prime}-l)}\phi_{q,k,m}\psi_{q,k,m^{\prime}}^{\ast
}(2n_{q,k}-1). \label{staticcontraction}%
\end{equation}
Therefore, in order to obtain the average spontaneous magnetization per cell,
$M^{x},$ we have to determine the asymptotic behaviour of the determinant
shown in eq.(\ref{determinante-sx}). For finite temperature, it can be shown
that the asymptotic behaviour of the determinant is zero, which corresponds to
the limit $r\rightarrow\infty,$ and, as expected, there is no spontaneous
magnetization at $T\neq0.$

\begin{figure}
[ptb]
\begin{center}
\includegraphics[width=0.5\columnwidth
]%
{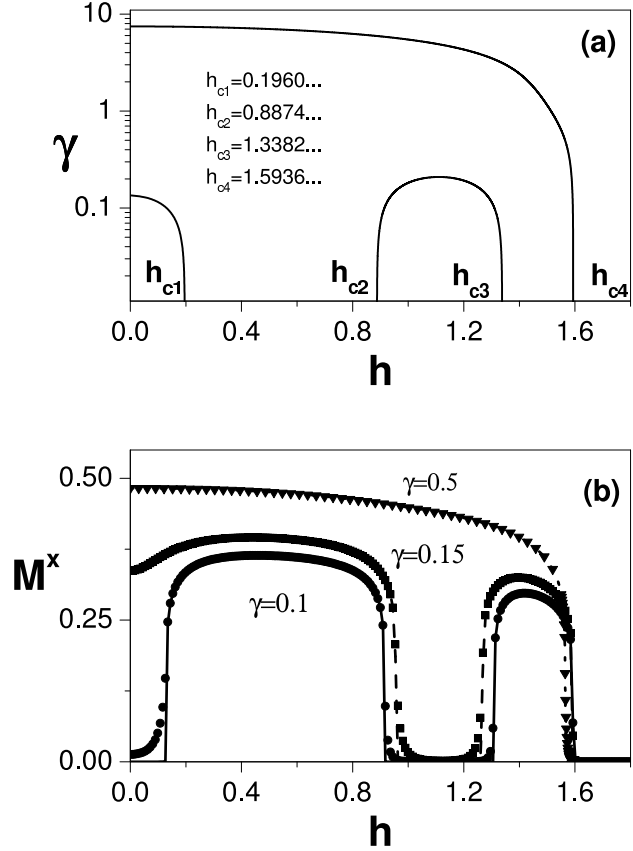}%
\caption{(a) Critical lines, at $T=0,$ for $n=4,$ $\mu_{1}=\mu_{2}=\mu_{3}%
=\mu_{4}=1$ and for $J_{1}=1,$ $J_{2}=1.5,$ $J_{3}=1.75,$ $J_{4}=2$. (b)
Spontaneous magnetization M$^{x}$ as a function of the uniform field, for
different $\gamma$, $n=4,$ $\mu_{1}=\mu_{2}=\mu_{3}=\mu_{4}=1$ and $J_{1}=1,$
$J_{2}=1.5,$ $J_{3}=1.75,$ $J_{4}=2$. The continuous line represents the
extrapolation procedure. }%
\label{fig6}%
\end{center}
\end{figure}

\begin{figure}
[ptb]
\begin{center}
\includegraphics[width=0.5\columnwidth
]%
{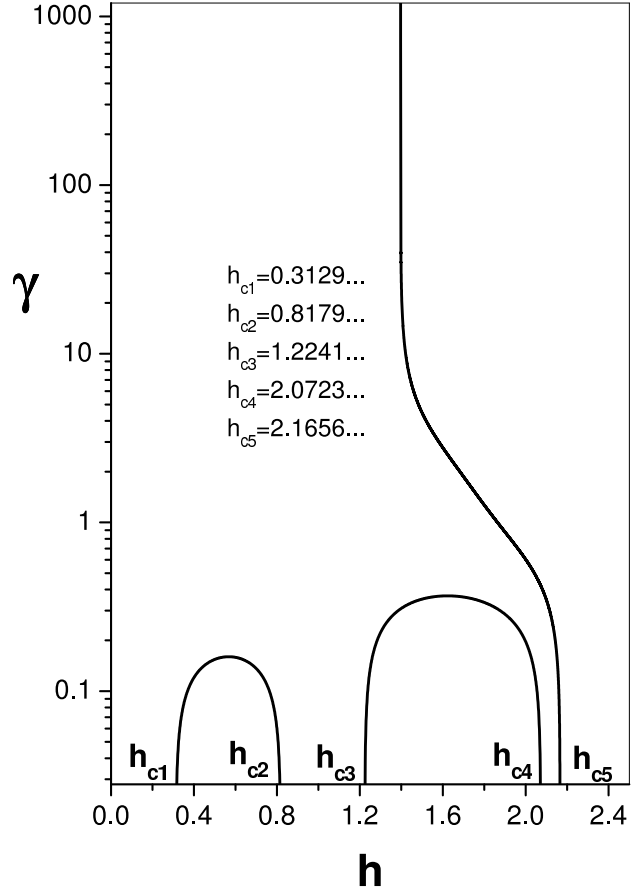}%
\caption{ Critical lines, at $T=0,$ for $n=5,$ $\mu_{1}=\mu_{2}=\mu_{3}%
=\mu_{4}=\mu_{5}=1$ and for $J_{1}=1,$ $J_{2}=1.5,$ $J_{3}=2,$ $J_{4}=2.5,$
$J_{5}=3$, where the vertical scale is decimal logarithmic.}%
\label{fig7}%
\end{center}
\end{figure}

On the other hand, at $T=0,$  the asymptotic behaviour of the determinant
is different from zero for values of the field
where the ground state is ordered. This asymptotic behaviour of the
determinant can be estimated numerically by using the $\varepsilon-$algorithm
\cite{barber:1983}, and, by using this method, we have obtained $M^{x}$ for a
lattice with $n=4,$ identical $\mu^{\prime}s$ and $\gamma^{\prime}s.$ In order
to obtain the estimate of the asymptotic value of the determinant, we have
considered a numerical series constructed by varying $r$ from 56 to 80, and
the results are shown in Fig.(\ref{fig6}b). The behaviour of the magnetization
close to the transition points has been obtained by adjusting the reliable
numerical results in the critical region to the scaling function%

\begin{equation}
M^{x}(h)=M_{0}\left\vert h-h_{c}\right\vert ^{\overline{\beta}}.
\label{ajusteMx}%
\end{equation}

\noindent By considering a non-linear regression we have been able to obtain
the adjustable parameters of the scaling function, namely, the critical field
$h_{c},$ the critical exponent $\overline{\beta}$ and the amplitude $M_{0}.$
These results are shown in Table I.

\begin{table}[p]
\begin{center}%
\begin{tabular}
[c]{c|c|c|c|c}\hline
$\gamma$ & $h_{c}$ (exact) & $M_{0}$ & $h_{c}$ (calculated) & $\overline
{\beta}$\\\hline
& $h_{c1}=0.130554559...$ & 0.479(8) & 0.13050(1) & 0.125(3)\\
$\gamma=0.1$ & $h_{c2}=0.914163156...$ & 0.465(4) & 0.9142(1) & 0.125(2)\\
& $h_{c3}=1.310209705...{\ }{\ }{\ }{\ }$ & 0.436(8) & 1.31017(6) & 0.125(4)\\
& $h_{c4}=1.5922623078...$ & 0.395(2) & 1.59227(6) & 0.126(1)\\\hline
& $h_{c1}=0.9541874918...$ & 0.488(12) & 0.9542(3) & 0.125(6)\\
$\gamma=0.15$ & $h_{c2}=1.2687172332...$ & 0.465(8) & 1.26870(5) & 0.125(4)\\
& $h_{c3}=1.5905738964..$ & 0.423(5) & 1.5906(1) & 0.125(3)\\\hline
$\gamma=0.5$ & $h_{c}=1.5644253446...$ & 0.501(9) & 1.5644(2) &
0.125(4)\\\hline
\end{tabular}
\end{center}
\caption{Numerical results for the fitting, in the scaling region, of the
spontaneous magnetization $M^{x}$ at $T=0,$ as a function of the uniform field
for $n=4,$ $\mu_{1}=\mu_{2}=\mu_{3}=\mu_{4}=1$, $\gamma_{1}=\gamma_{2}%
=\gamma_{3}=\gamma_{4}=\gamma,$ $J_{1}=1$, $J_{2}=1.5$, $J_{3}=1.75,$
$J_{4}=2$ and different $\gamma.$ The parameters shown have been obtained from
the non-linear regression of the equation $M^{x}(h)=M_{0}\left\vert
h_{c}-h\right\vert ^{\overline{\beta}}$}%
\label{table}%
\end{table}

As it can be seen in Table I, the comparison of the numerical results,
obtained from the non-linear regression, with the exact known ones show that
they are extremely precise. This rather surprising result means that the
quantum critical region can be precisely described numerically by
eq.(\ref{ajusteMx}).%

%

\bigskip

\section{Static and dynamic correlations $\left\langle \tau_{l}^{z}\tau
_{l+r}^{z}\right\rangle $}

The correlation function $<S_{l}^{z}(t)S_{l+r}^{z}(0)>,$ in the thermodynamic
limit, can be obtained from the expression
\cite{siskens:1974,capel:1977,goncalvestese:1977},
\begin{equation}
\left\langle S_{l,m}^{z}(t)S_{l+r,m^{\prime}}^{z}(0)\right\rangle
=\frac{Tr\left[  \exp(-\beta H^{-})\exp(iH^{-}t)S_{l,m}^{z}\exp(-iH^{-}%
t)S_{l+r,m^{\prime}}^{z}\right]  }{Tr\left[  \exp(-\beta H^{-})\right]  }.
\end{equation}
Therefore, the dynamic correlation between the effective spins in the field
direction,%
\begin{equation}
\left\langle \tau_{l}^{z}(t)\tau_{l+r}^{z}(0)\right\rangle =\frac{1}{n^{2}%
}\sum_{m,m^{\prime}=1}^{n}\mu_{m}\mu_{m^{\prime}}\left\langle S_{l,m}%
^{z}(t)S_{l+r,m^{\prime}}^{z}(0)\right\rangle ,
\end{equation}
by using eqs. (\ref{Jordan-Wigner}) and (\ref{comblinferm}), and Wick's
theorem, can be expressed as
\begin{align}
\left\langle \tau_{l}^{z}(t)\tau_{l+r}^{z}(0)\right\rangle  &  =\frac
{1}{4n^{2}}\sum_{m=1}^{n}\sum_{m^{\prime}=1}^{n}\mu_{m}\mu_{m^{\prime}%
}\left\{  \left\langle A_{l,m}(t)B_{l,m}(t)\right\rangle \left\langle
A_{l+r,m^{\prime}}(0)B_{l+r,m^{\prime}}(0)\right\rangle \right. \nonumber-\\
&  -\left\langle A_{l,m}(t)A_{l+r,m^{\prime}}(0)\right\rangle \left\langle
B_{l,m}(t)B_{l+r,m^{\prime}}(0)\right\rangle \nonumber+\\
&  +\left.  \left\langle A_{l,m}(t)B_{l+r,m^{\prime}}(0)\right\rangle
\left\langle B_{l,m}(t)A_{l+r,m^{\prime}}(0)\right\rangle \right\}  ,
\label{dynCorrWick}%
\end{align}

where the dynamic contractions, obtained from eq.(\ref{AeB}), are given by%

\begin{align}
\left\langle A_{l,m}(t)A_{l+r,m^{\prime}}(0)\right\rangle  &  =\frac{1}{N}%
\sum_{q,k}e^{-iqdr}\psi_{q,k,m}\psi_{q,k,m^{\prime}}^{\ast}\left[
e^{iE_{q,k}t}n_{q,k}+e^{-iE_{q,k}t}\left(  1-n_{q,k}\right)  \right]
,\nonumber\\
\left\langle B_{l,m}(t)B_{l+r,m^{\prime}}(0)\right\rangle  &  =\frac{1}{N}%
\sum_{q,k}e^{-iqdr}\phi_{q,k,m}\phi_{q,k,m^{\prime}}^{\ast}\left[
-\ e^{iE_{q,k}t}n_{q,k}-e^{-iE_{q,k}t}\left(  1-n_{q,k}\right)  \right]
,\nonumber\\
\left\langle A_{l,m}(t)B_{l+r,m^{\prime}}(0)\right\rangle  &  =\frac{1}{N}%
\sum_{q,k}e^{-iqdr}\psi_{q,k,m}\phi_{q,k,m^{\prime}}^{\ast}\left[
-\ e^{iE_{q,k}t}n_{q,k}+e^{-iE_{q,k}t}\left(  1-n_{q,k}\right)  \right]
,\nonumber\\
\left\langle B_{l,m}(t)A_{l+r,m^{\prime}}(0)\right\rangle  &  =\frac{1}{N}%
\sum_{q,k}e^{-iqdr}\phi_{q,k,m}\psi_{q,k,m^{\prime}}^{\ast}\left[
e^{iE_{q,k}t}n_{q,k}-e^{-iE_{q,k}t}\left(  1-n_{q,k}\right)  \right]  .
\label{contractions}%
\end{align}

Then we can write the dynamic correlation $\left\langle \tau_{l}^{z}%
(t)\tau_{l+r}^{z}(0)\right\rangle $ in the form%
\begin{align}
&  \left\langle \tau_{l}^{z}(t)\tau_{l+r}^{z}(0)\right\rangle =\left\langle
\tau_{l}^{z}\right\rangle ^{2}+\nonumber\\
&  +\frac{1}{4n^{2}N^{2}}\sum_{q,k,m}\sum_{q^{\prime},k^{\prime},m^{\prime}%
}\mu_{m}\mu_{m^{\prime}}e^{-i(q-q^{\prime})dr}\psi_{q,k,m}\psi_{q,k,m^{\prime
}}^{\ast}\phi_{q^{\prime},k^{\prime},m^{\prime}}\phi_{q^{\prime},k^{\prime}%
,m}^{\ast}\times\nonumber\\
&  \times\left[  e^{i(E_{q,k}+E_{q^{\prime},k^{\prime}})t}n_{q,k}n_{q^{\prime
},k^{\prime}}+e^{-i(E_{q,k}+E_{q^{\prime},k^{\prime}})t}\left(  1-n_{q,k}%
\right)  \left(  1-n_{q^{\prime},k^{\prime}}\right)  \right.  +\nonumber\\
&  +\left.  e^{i(E_{q,k}-E_{q^{\prime},k^{\prime}})t}n_{q,k}\left(
1-n_{q^{\prime},k^{\prime}}\right)  +e^{-i(E_{q,k}-E_{q^{\prime},k^{\prime}%
})t}\left(  1-n_{q,k}\right)  n_{q^{\prime},k^{\prime}}\right]  -\nonumber\\
&  -\frac{1}{4n^{2}N^{2}}\sum_{q,k,m}\sum_{q^{\prime},k^{\prime},m^{\prime}%
}\mu_{m}\mu_{m^{\prime}}e^{-i(q-q^{\prime})dr}\psi_{q,k,m}\phi_{q,k,m^{\prime
}}^{\ast}\psi_{q^{\prime},k^{\prime},m^{\prime}}\phi_{q^{\prime},k^{\prime}%
,m}^{\ast}\times\nonumber\\
&  \times\left[  e^{i(E_{q,k}+E_{q^{\prime},k^{\prime}})t}n_{q,k}n_{q^{\prime
},k^{\prime}}+e^{-i(E_{q,k}+E_{q^{\prime},k^{\prime}})t}\left(  1-n_{q,k}%
\right)  \left(  1-n_{q^{\prime},k^{\prime}}\right)  \right.  -\nonumber\\
&  -\left.  e^{i(E_{q,k}-E_{q^{\prime},k^{\prime}})t}n_{q,k}\left(
1-n_{q^{\prime},k^{\prime}}\right)  -e^{-i(E_{q,k}-E_{q^{\prime},k^{\prime}%
})t}\left(  1-n_{q,k}\right)  n_{q^{\prime},k^{\prime}}\right]
.\label{dynamiccorrzz}%
\end{align}
For $t=0$ we obtain the static correlation, which is given by%

\begin{align}
&  \left\langle \tau_{l}^{z}(0)\tau_{l+r}^{z}(0)\right\rangle =\left\langle
\tau_{l}^{z}\right\rangle ^{2}+\nonumber\\
&  +\frac{1}{4n^{2}N^{2}}\sum_{q,k,m}\sum_{q^{\prime},k^{\prime},m^{\prime}%
}\mu_{m}\mu_{m^{\prime}}e^{-i(q-q^{\prime})dr}\psi_{q,k,m}\psi_{q,k,m^{\prime
}}^{\ast}\phi_{q^{\prime},k^{\prime},m^{\prime}}\phi_{q^{\prime},k^{\prime}%
,m}^{\ast}\nonumber-\\
&  -\frac{1}{4n^{2}N^{2}}\sum_{q,k,m}\sum_{q^{\prime},k^{\prime},m^{\prime}%
}\mu_{m}\mu_{m^{\prime}}e^{-i(q-q^{\prime})dr}\psi_{q,k,m}\phi_{q,k,m^{\prime
}}^{\ast}\psi_{q^{\prime},k^{\prime},m^{\prime}}\phi_{q^{\prime},k^{\prime}%
,m}^{\ast}\times\nonumber\\
&  \times\left(  1-2n_{q,k}\right)  \left(  1-2n_{q^{\prime},k^{\prime}%
}\right)  . \label{staticcorrzz}%
\end{align}

At $T=0$, the dynamic and static correlation are given respectively by%

\begin{align}
&  \left\langle \tau_{l}^{z}(t)\tau_{l+r}^{z}(0)\right\rangle =\left\langle
\tau_{l}^{z}\right\rangle ^{2}+\nonumber\\
&  +\frac{1}{4n^{2}N^{2}}\sum_{q,k,m}\sum_{q^{\prime},k^{\prime},m^{\prime}%
}\mu_{m}\mu_{m^{\prime}}e^{-i(q-q^{\prime})dr}\psi_{q,k,m}\psi_{q,k,m^{\prime
}}^{\ast}\phi_{q^{\prime},k^{\prime},m^{\prime}}\phi_{q^{\prime},k^{\prime}%
,m}^{\ast}\times\nonumber\\
&  \times\left[  e^{-i(E_{q,k}+E_{q^{\prime},k^{\prime}})t}\right]
-\nonumber\\
&  -\frac{1}{4n^{2}N^{2}}\sum_{q,k,m}\sum_{q^{\prime},k^{\prime},m^{\prime}%
}\mu_{m}\mu_{m^{\prime}}e^{-i(q-q^{\prime})dr}\psi_{q,k,m}\phi_{q,k,m^{\prime
}}^{\ast}\psi_{q^{\prime},k^{\prime},m^{\prime}}\phi_{q^{\prime},k^{\prime}%
,m}^{\ast}\times\nonumber\\
&  \times\left[  e^{-i(E_{q,k}+E_{q^{\prime},k^{\prime}})t}\right]
\label{dynamiccorrzzT=0}%
\end{align}

and%
\begin{align}
&  \left\langle \tau_{l}^{z}(0)\tau_{l+r}^{z}(0)\right\rangle =\left\langle
\tau_{l}^{z}\right\rangle ^{2}+\nonumber\\
&  +\frac{1}{4n^{2}N^{2}}\sum_{q,k,m}\sum_{q^{\prime},k^{\prime},m^{\prime}%
}\mu_{m}\mu_{m^{\prime}}e^{-i(q-q^{\prime})dr}\psi_{q,k,m}\psi_{q,k,m^{\prime
}}^{\ast}\phi_{q^{\prime},k^{\prime},m^{\prime}}\phi_{q^{\prime},k^{\prime}%
,m}^{\ast}\nonumber-\\
&  -\frac{1}{4n^{2}N^{2}}\sum_{q,k,m}\sum_{q^{\prime},k^{\prime},m^{\prime}%
}\mu_{m}\mu_{m^{\prime}}e^{-i(q-q^{\prime})dr}\psi_{q,k,m}\phi_{q,k,m^{\prime
}}^{\ast}\psi_{q^{\prime},k^{\prime},m^{\prime}}\phi_{q^{\prime},k^{\prime}%
,m}^{\ast}. \label{staticcoozzT=0}%
\end{align}

%

\begin{figure}
[ptb]
\begin{center}
\includegraphics[width=0.5\columnwidth
]%
{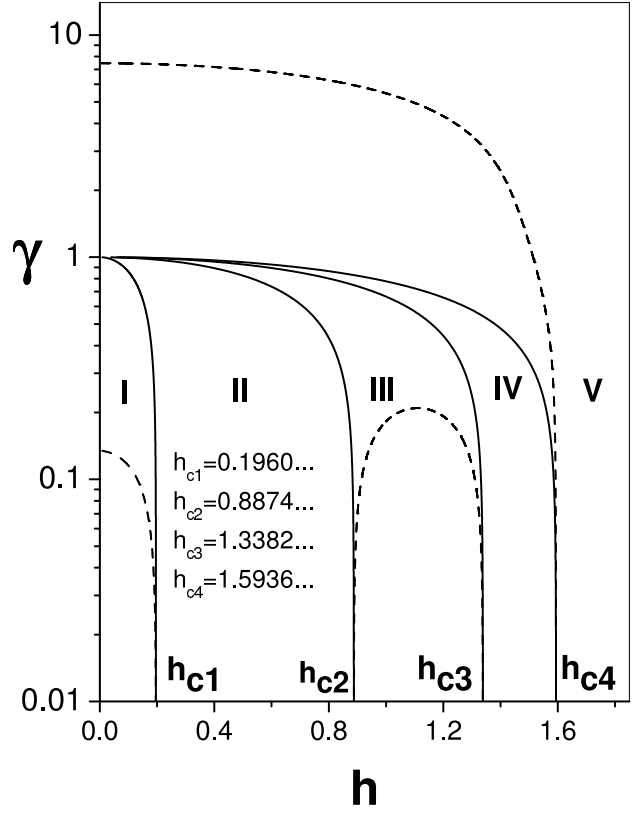}%
\caption{Disorder lines (continuous line) and critical (dashed line), at
$T=0,$ for $n=4,$ $\mu_{1}=\mu_{2}=\mu_{3}=\mu_{4}=1$ and for $J_{1}=1,$
$J_{2}=1.5,$ $J_{3}=1.75,$ $J_{4}=2$, where the vertical scale is decimal logarithmic. The asymptotic behaviour alternates
between monotonic and oscillatory as we go from region I to region V.}%
\label{fig8}%
\end{center}
\end{figure}

\bigskip%
\begin{figure}
[ptb]
\begin{center}
\includegraphics[width=0.5\columnwidth
]%
{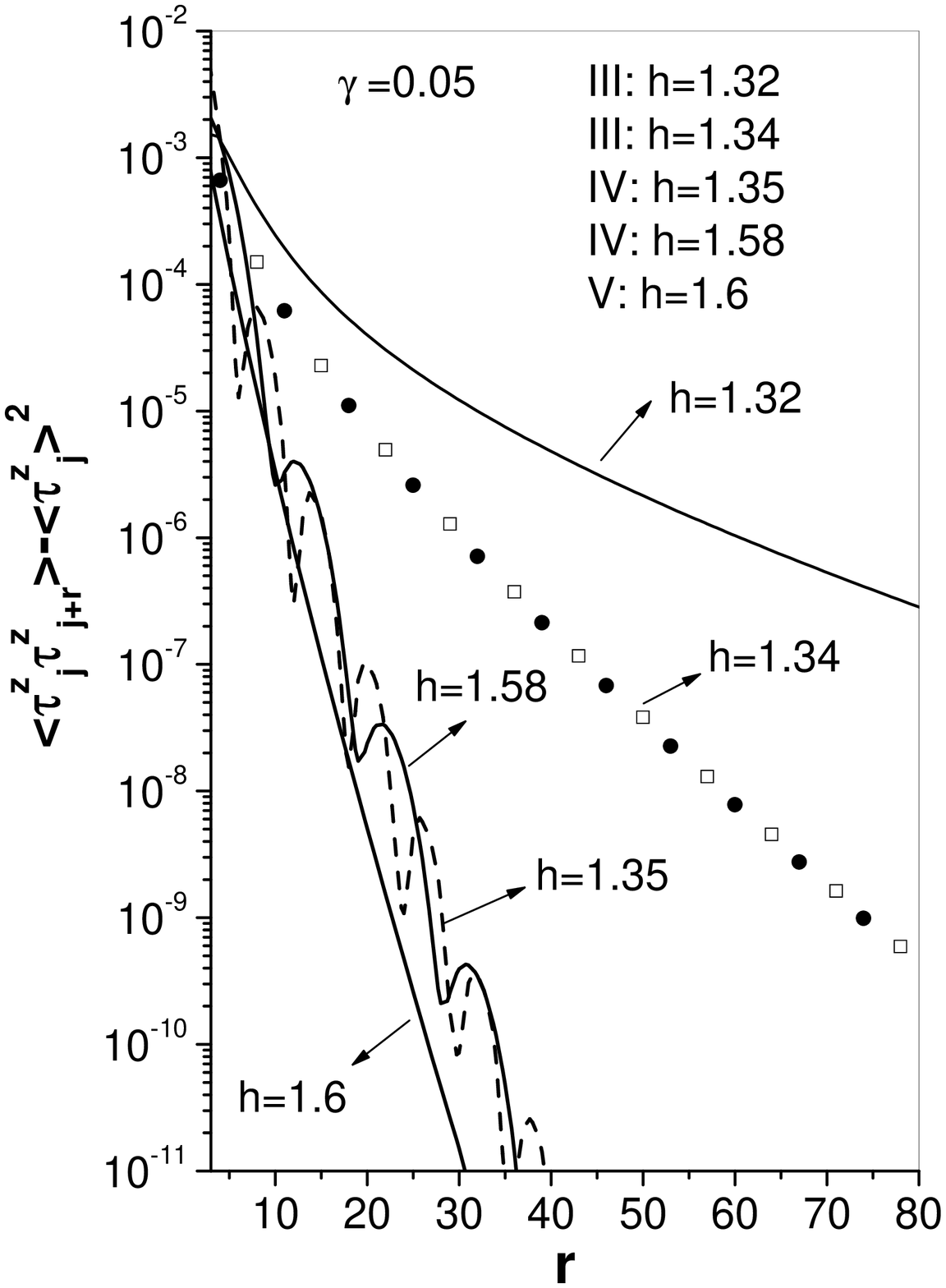}%
\caption{Static correlation function in the field direction$,$ at $T=0,$ for
$n=4,$ $\mu_{1}=\mu_{2}=\mu_{3}=\mu_{4}=1$ and for $J_{1}=1,$ $J_{2}=1.5,$
$J_{3}=1.75,$ $J_{4}=2$, as function of $r$ (distance between cells) for
values of the field belonging to different regions shown in Fig. 8, where the vertical scale is decimal logarithmic.}%
\label{fig9}%
\end{center}
\end{figure}
\begin{figure}
[ptb]
\begin{center}
\includegraphics[width=0.5\columnwidth
]%
{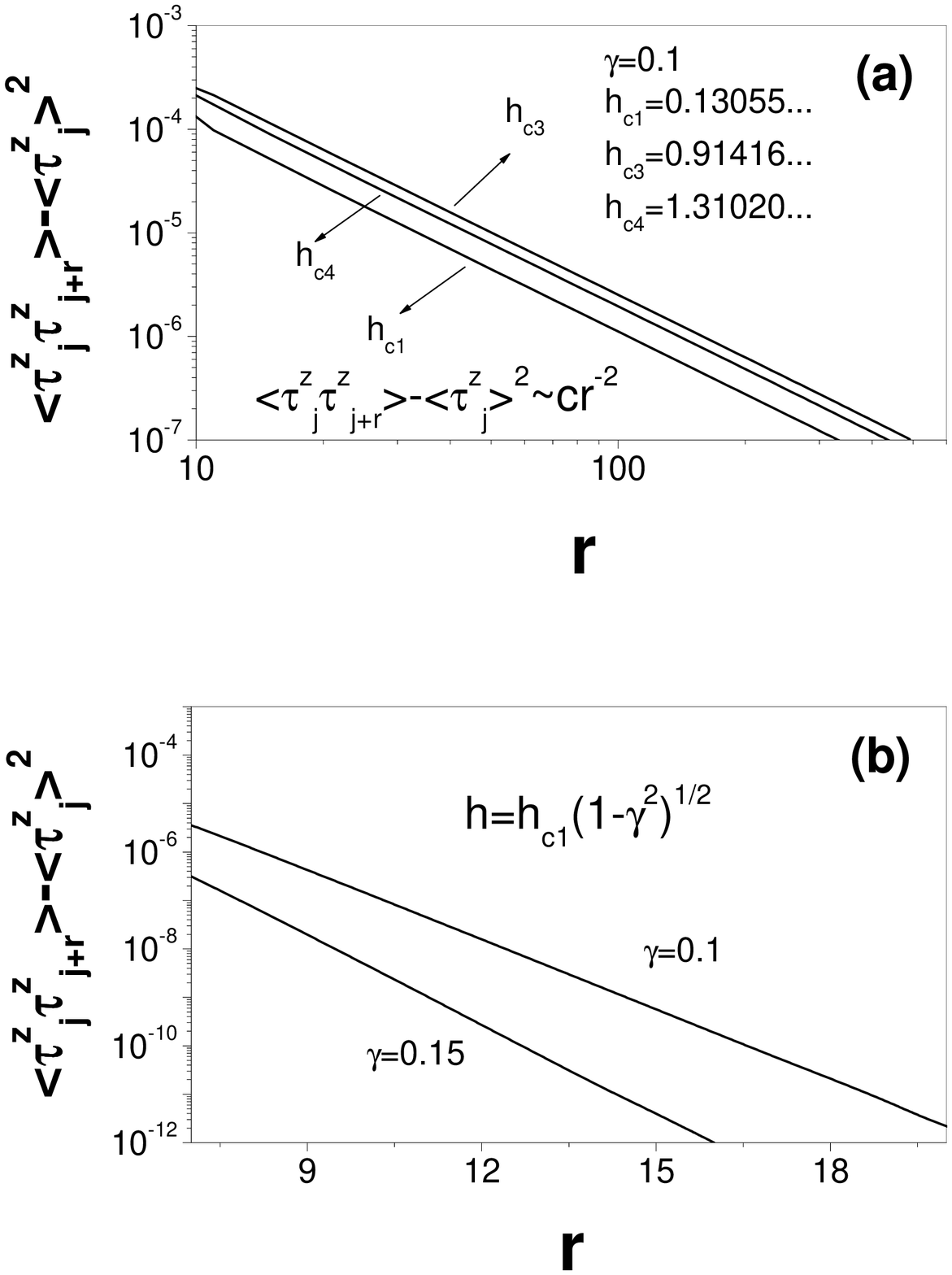}%
\caption{Asymptotic behaviour of the direct static correlation function in the
field direction$,$ at $T=0,$ for $n=4,$ $\mu_{1}=\mu_{2}=\mu_{3}=\mu_{4}=1$
and for $J_{1}=1,$ $J_{2}=1.5,$ $J_{3}=1.75,$ $J_{4}=2$, as function of $r$
(distance between cells) for values of the field belonging to the critical
lines (a), where the scales are decimal logarithmic, and disorder lines (b), where the vertical scale is also decimal logarithmic.}%
\label{fig10}%
\end{center}
\end{figure}
As in the homogenous model \cite{barouch:1971}, the asymptotic behaviour of
the static correlation $\left\langle \tau_{l}^{z}(0)\tau_{l+r}^{z}%
(0)\right\rangle $, at $T=0$, as a function of $r,$ can be oscillatory or
monotonic depending on the values of the interaction parameters. For arbitrary
$n$, these regions are separated by hypersurfaces in the parameter space and
are called disordered surfaces, which collapse into disorder lines for
identical $\gamma^{\prime}s$.\ For $n=4,$ these disorder lines and the
critical lines are shown in Fig.(\ref{fig8}). Independently of the number of
sites in unit cell, in the plane $\gamma\times h,$ the disorder lines are
given by equation%
\begin{equation}
h^{2}+\gamma^{2}h_{jc}^{2}=h_{jc}^{2}, \label{disorderedline}%
\end{equation}
where $h_{jc}$ is a critical field. Along these curves the anisotropic model
is equivalent to an effective isotropic model \cite{barouch:1971}. For $n=2$,
this equivalence is shown in Appendix A, and for $n>2$ it has been verified
numerically$.$ It is also shown, that, along this curve, the effective
isotropic model is in a disordered state, which corresponds to the
magnetization plateaus for the case where we have identical $\mu^{\prime}$s.

The asymptotic behaviour of the static correlation function, also for $n=4$
and at $T=0,$ as a function of the distance $r$ between cells, is presented in
Fig.(\ref{fig9}) for various values of $h$ and $\gamma$ in the different
regions shown in Fig.(\ref{fig8}). As can be seen, the behaviour, as expected
alternates between oscillatory and monotonic.

At the critical point, as in the homogeneous case, the direct static
correlation behaves asymptotically as $r^{-2}$\cite{barouch:1971}. This
behaviour is shown in Fig.(\ref{fig10}a), for $n=4$, and different critical
fields. However, at the disorder line, differently from the homogeneous model
where the direct static correlation is zero, it behaves asymptotically as
$\lambda^{-r},$ where $\lambda$ is a function of the field and of the
interaction parameters, as shown in Fig.(\ref{fig10}b), for identical
$\gamma^{\prime}s$ and different values of the field.

\begin{figure}
[ptb]
\begin{center}
\includegraphics[width=0.5\columnwidth
]%
{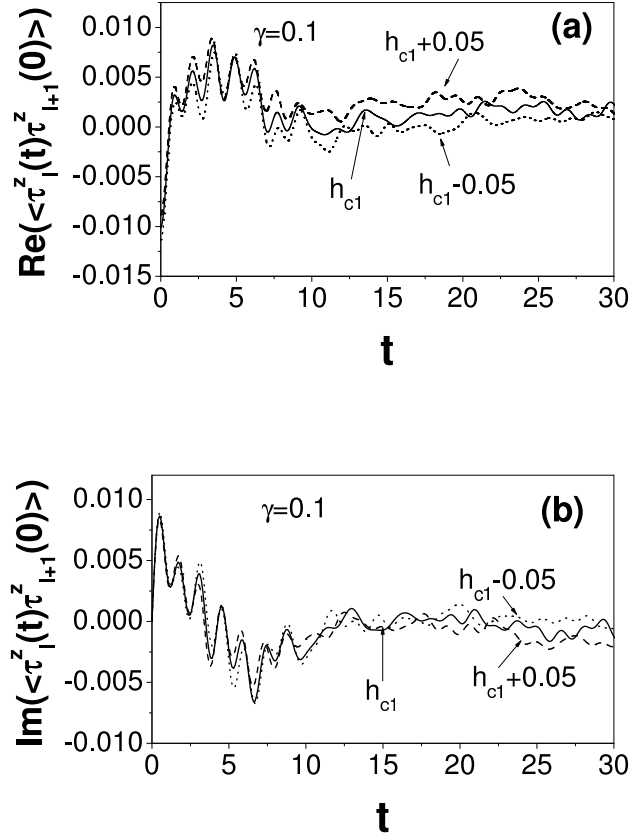}%
\caption{(a) The real and (b) imaginary parts of the dynamic correlation
function $\left\langle \tau_{l}^{z}(t)\tau_{l+1}^{z}(0)\right\rangle ,$ at
$T=0$ for $n=4,$ $\mu_{1}=\mu_{2}=\mu_{3}=\mu_{4}=1$ and $J_{1}=1,$
$J_{2}=1.5,$ $J_{3}=1.75,$ $J_{4}=2,$ as function of time $t.$}%
\label{fig11}%
\end{center}
\end{figure}

The real and imaginary parts of the dynamic correlation for a chain with $n=4$
and uniform $\mu^{\prime}s,$ at $T=0,$ are presented in Fig.(\ref{fig11}) for
values of the field greater and equal to the critical field. As in the
isotropic case, apart from the asymptotic behaviour, no noticeable difference
is observed in the dynamic correlation at different values of the field.

\section{Dynamic Susceptibility}

From the dynamic correlation $\left\langle \tau_{l}^{z}(t)\tau_{l+r}%
^{z}(0)\right\rangle ,$ eqs.(\ref{dynamiccorrzz}), we can obtain the time
Fourier transform of dynamic correlation $\left\langle \tau_{l}^{z}\tau
_{l+r}^{z}\right\rangle _{\omega}$ from the equation%
\begin{equation}
\left\langle \tau_{l}^{z}\tau_{l+r}^{z}\right\rangle _{\omega}=\frac{1}{2\pi}%
{\displaystyle\int\limits_{-\infty}^{\infty}}
\left\langle \tau_{l}^{z}(t)\tau_{l+r}^{z}(0)\right\rangle e^{i\omega t}dt,
\label{Time FourierTransform}%
\end{equation}
which is given by%
\begin{align}
\left\langle \tau_{l}^{z}\tau_{l+r}^{z}\right\rangle _{\omega}  &
=\delta(\omega)\left\langle \tau_{l}^{z}\right\rangle ^{2}+\nonumber\\
&  +\frac{1}{4n^{2}N^{2}}\sum_{q,k,m}\sum_{q^{\prime},k^{\prime},m^{\prime}%
}\mu_{m}\mu_{m^{\prime}}e^{-i(q-q^{\prime})dr}\psi_{q,k,m}\psi_{q,k,m^{\prime
}}^{\ast}\phi_{q^{\prime},k^{\prime},m^{\prime}}\phi_{q^{\prime},k^{\prime}%
,m}^{\ast}\times\nonumber\\
&  \times\left[  \delta\left(  \omega+E_{q,k}+E_{q^{\prime},k^{\prime}%
}\right)  n_{q,k}n_{q^{\prime},k^{\prime}}\right.  +\left.  \delta\left(
\omega-E_{q,k}-E_{q^{\prime},k^{\prime}}\right)  \left(  1-n_{q,k}\right)
\left(  1-n_{q^{\prime},k^{\prime}}\right)  \right.  +\nonumber\\
&  +\left.  \delta\left(  \omega+E_{q,k}-E_{q^{\prime},k^{\prime}}\right)
n_{q,k}\left(  1-n_{q^{\prime},k^{\prime}}\right)  \right.  +\left.
\delta\left(  \omega-E_{q,k}+E_{q^{\prime},k^{\prime}}\right)  \left(
1-n_{q,k}\right)  n_{q^{\prime},k^{\prime}}\right] \nonumber-\\
&  -\frac{1}{4n^{2}N^{2}}\sum_{q,k,m}\sum_{q^{\prime},k^{\prime},m^{\prime}%
}\mu_{m}\mu_{m^{\prime}}e^{-i(q-q^{\prime})dr}\psi_{q,k,m}\phi_{q,k,m^{\prime
}}^{\ast}\psi_{q^{\prime},k^{\prime},m^{\prime}}\phi_{q^{\prime},k^{\prime}%
,m}^{\ast}\times\nonumber\\
&  \times\left[  \delta\left(  \omega+E_{q,k}+E_{q^{\prime},k^{\prime}%
}\right)  n_{q,k}n_{q^{\prime},k^{\prime}}+\delta\left(  \omega-E_{q,k}%
-E_{q^{\prime},k^{\prime}}\right)  \left(  1-n_{q,k}\right)  \left(
1-n_{q^{\prime},k^{\prime}}\right)  -\right. \nonumber\\
-  &  \left.  \delta\left(  \omega+E_{q,k}-E_{q^{\prime},k^{\prime}}\right)
n_{q,k}\left(  1-n_{q^{\prime},k^{\prime}}\right)  -\delta\left(
\omega-E_{q,k}+E_{q^{\prime},k^{\prime}}\right)  \left(  1-n_{q,k}\right)
n_{q^{\prime},k^{\prime}}\right]  . \label{FTcorrzz}%
\end{align}
By introducing the spatial Fourier transform%
\begin{equation}
\left\langle \tau_{q}^{z}\tau_{-q}^{z}\right\rangle _{\omega}=%
{\displaystyle\sum\limits_{r}}
\left\langle \tau_{l}^{z}\tau_{l+r}^{z}\right\rangle _{\omega}e^{idrq},
\label{spatialFTcorrzz}%
\end{equation}
in the previous expression, we can write immediately%
\begin{align}
\left\langle \tau_{q}^{z}\tau_{-q}^{z}\right\rangle _{\omega}  &
=N\delta_{q,0}\delta(\omega)\left\langle \tau_{l}^{z}\right\rangle ^{2}+\nonumber\\
&  +\frac{1}{4n^{2}N}\sum_{q^{\prime},k,m,k^{\prime},m^{\prime}}\mu_{m}%
\mu_{m^{\prime}}\psi_{q^{\prime},k,m}\psi_{q^{\prime},k,m^{\prime}}^{\ast}%
\phi_{q^{\prime}-q,k^{\prime},m^{\prime}}\phi_{q^{\prime}-q,k^{\prime}%
,m}^{\ast}\times\nonumber\\
&  \times\left[  \delta\left(  \omega+E_{q^{\prime},k}+E_{q^{\prime
}-q,k^{\prime}}\right)  n_{q^{\prime},k}n_{q^{\prime}-q,k^{\prime}}\right.
+\left.  \delta\left(  \omega-E_{q^{\prime},k}-E_{q^{\prime}-q,k^{\prime}%
}\right)  \left(  1-n_{q^{\prime},k}\right)  \left(  1-n_{q^{\prime
}-q,k^{\prime}}\right)  \right.  +\nonumber\\
&  +\left.  \delta\left(  \omega+E_{q^{\prime},k}-E_{q^{\prime}-q,k^{\prime}%
}\right)  n_{q^{\prime},k}\left(  1-n_{q^{\prime}-q,k^{\prime}}\right)
\right.  +\left.  \delta\left(  \omega-E_{q^{\prime},k}+E_{q^{\prime
}-q,k^{\prime}}\right)  \left(  1-n_{q^{\prime},k}\right)  n_{q^{\prime
}-q,k^{\prime}}\right] \nonumber-\\
&  -\frac{1}{4n^{2}N}\sum_{q^{\prime},k,m,k^{\prime},m^{\prime}}\mu_{m}%
\mu_{m^{\prime}}\psi_{q^{\prime},k,m}\phi_{q^{\prime},k,m^{\prime}}^{\ast}%
\psi_{q^{\prime}-q,k^{\prime},m^{\prime}}\phi_{q^{\prime}-q,k^{\prime}%
,m}^{\ast}\times\nonumber\\
&  \times\left[  \delta\left(  \omega+E_{q^{\prime},k}+E_{q^{\prime
}-q,k^{\prime}}\right)  n_{q^{\prime},k}n_{q^{\prime}-q,k^{\prime}}\right.
+\left.  \delta\left(  \omega-E_{q^{\prime},k}-E_{q-q^{\prime},k^{\prime}}\right)  \left(
1-n_{q^{\prime},k}\right)  \left(  1-n_{q-q^{\prime},k^{\prime}}\right)  \right.  -\nonumber\\
&  -\left.  \delta\left(  \omega+E_{q^{\prime},k}-E_{q^{\prime}-q,k^{\prime}%
}\right)  n_{q^{\prime},k}\left(  1-n_{q^{\prime}-q,k^{\prime}}\right)
\right.  -\left.  \delta\left(  \omega-E_{q^{\prime},k}+E_{q^{\prime
}-q,k^{\prime}}\right)  \left(  1-n_{q^{\prime},k}\right)  n_{q^{\prime
}-q,k^{\prime}}\right]  . \label{spacialtimeFTcorrzz}%
\end{align}

The dynamic susceptibility $\chi_{q}^{zz}(\omega)$ can be obtained by using
the expression \cite{zubarev:1960}%

\begin{equation}
\chi_{q}^{zz}(\omega)=-%
{\displaystyle\int\limits_{\infty}^{\infty}}
\frac{(1-e^{-\omega^{\prime}/k_{B}T})\left\langle \tau_{q}^{z}\tau_{-q}%
^{z}\right\rangle _{\omega^{\prime}}d\omega^{\prime}}{\omega-\omega^{\prime}},
\label{chizzomegaq}%
\end{equation}

and from this we obtain%
\begin{align}
\chi_{q}^{zz}(\omega)  &  =-\frac{1}{4n^{2}N}\sum_{q^{\prime},k,m,k^{\prime
},m^{\prime}}\mu_{m}\mu_{m^{\prime}}\left(  P_{q^{\prime},k,m,q^{\prime
}-q,k^{\prime},m^{\prime}}-Q_{q^{\prime},k,m,q^{\prime}-q,k^{\prime}%
,m^{\prime}}\right)  \times\nonumber\\
&  \times\left(  \frac{1}{\omega+E_{q^{\prime},k}+E_{q^{\prime}-q,k^{\prime}}%
}-\frac{1}{\omega-E_{q^{\prime},k}-E_{q^{\prime}-q,k^{\prime}}}\right)
\times\left[  n_{q^{\prime},k}+n_{q^{\prime}-q,k^{\prime}}-1\right]
-\nonumber\\
&  -\frac{1}{4n^{2}N}\sum_{q^{\prime},k,m,k^{\prime},m^{\prime}}\mu_{m}%
\mu_{m^{\prime}}\left(  P_{q^{\prime},k,m,q^{\prime}-q,k^{\prime},m^{\prime}%
}+Q_{q^{\prime},k,m,q^{\prime}-q,k^{\prime},m^{\prime}}\right)  \times
\nonumber\\
&  \times\left(  \frac{1}{\omega+E_{q^{\prime},k}-E_{q^{\prime}-q,k^{\prime}}%
}-\frac{1}{\omega-E_{q^{\prime},k}+E_{q^{\prime}-q,k^{\prime}}}\right)
\times\left[  n_{q^{\prime},k}-n_{q^{\prime}-q,k^{\prime}}\right]  ,
\label{susceptibilidade_dinamica-aniso}%
\end{align}
where
\begin{align}
P_{q^{\prime},k,m,q^{\prime}-q,k^{\prime},m^{\prime}}  &  =\psi_{q^{\prime
},k,m}\psi_{q^{\prime},k,m^{\prime}}^{\ast}\phi_{q^{\prime}-q,k^{\prime
},m^{\prime}}\phi_{q^{\prime}-q,,k^{\prime},m}^{\ast},\label{PQaniso}\\
Q_{q^{\prime},k,m,q^{\prime}-q,k^{\prime},m^{\prime}}  &  =\psi_{q^{\prime
},k,m}\phi_{q^{\prime},k,m^{\prime}}^{\ast}\psi_{q^{\prime}-q,k^{\prime
},m^{\prime}}\phi_{q^{\prime}-q,k^{\prime},m}^{\ast},
\end{align}
and we have used the identity%
\begin{equation}
1-n_{q,k}=e^{\beta E_{q,k}}n_{q,k}.
\end{equation}
It should be noted that the previous result reduces to the known one for the
isotropic model\cite{delima:2006}.

The static susceptibility $\chi_{q}^{zz}(0)$ is obtained by making $\omega=0$
in eq.(\ref{susceptibilidade_dinamica-aniso}), and is given by
\begin{align}
\chi_{q}^{zz}(0)  &  =-\frac{1}{2n^{2}N}\sum_{q^{\prime},k,m,k^{\prime
},m^{\prime}}\mu_{m}\mu_{m^{\prime}}\left(  P_{q^{\prime},k,m,q^{\prime
}-q,k^{\prime},m^{\prime}}-Q_{q^{\prime},k,m,q^{\prime}-q,k^{\prime}%
,m^{\prime}}\right)  \times\nonumber\\
&  \times\left(  \frac{1}{E_{q^{\prime},k}+E_{q^{\prime}-q,k^{\prime}}%
}\right)  \times\left[  n_{q^{\prime},k}+n_{q^{\prime}-q,k^{\prime}}-1\right]
\nonumber-\\
&  -\frac{1}{2n^{2}N^{2}}\sum_{q^{\prime},k,m,k^{\prime},m^{\prime}}\left(
P_{q^{\prime},k,m,q^{\prime}-q,k^{\prime},m^{\prime}}+Q_{q^{\prime
},k,m,q^{\prime}-q,k^{\prime},m^{\prime}}\right)  \times\nonumber\\
&  \times\left(  \frac{1}{E_{q^{\prime},k}-E_{q^{\prime}-q,k^{\prime}}%
}\right)  \times\left[  n_{q^{\prime},k}-n_{q^{\prime}-q,k^{\prime}}\right]  .
\label{susceptibilidade_estatica-aniso-kubo}%
\end{align}

The isothermal susceptibility can also be obtained by using the dynamic
correlation given in eq.(\ref{dynamiccorrzz}), and can be written
as\cite{Marshall:1971}%
\begin{equation}
\chi_{T}^{zz}=%
{\displaystyle\sum\limits_{r}}
{\displaystyle\int\limits_{0}^{\beta}}
\left(  \left\langle \tau_{l}^{z}(-i\lambda)\tau_{l+r}^{z}(0)\right\rangle
-\left\langle \tau_{l}^{z}\right\rangle ^{2}\right)  d\lambda,
\label{isothermalKubo}%
\end{equation}
where, as defined previously, $\beta=1/k_{B}T.$ From this it can be shown that
the isothermal susceptibility in the field direction is equal to the uniform
static one $\chi_{0}^{zz}(0)$.

At $T=0$, the dynamic and static susceptibilities given in
eqs.(\ref{susceptibilidade_dinamica-aniso}) and
(\ref{susceptibilidade_estatica-aniso-kubo}) can be explicitly written as%
\begin{align}
\chi_{q}^{zz}(\omega)  &  =\frac{1}{4n^{2}N}\sum_{q^{\prime},k,m,k^{\prime
},m^{\prime}}\mu_{m}\mu_{m^{\prime}}\left(  P_{q^{\prime},k,m,q^{\prime
}-q,k^{\prime},m^{\prime}}-Q_{q^{\prime},k,m,q^{\prime}-q,k^{\prime}%
,m^{\prime}}\right)  \times\\
&  \times\left(  \frac{1}{\omega+E_{q^{\prime},k}+E_{q^{\prime}-q,k^{\prime}}%
}-\frac{1}{\omega-E_{q^{\prime},k}-E_{q^{\prime}-q,k^{\prime}}}\right)  ,
\label{susdynamicsT=0}%
\end{align}
and%
\begin{align}
\chi_{q}^{zz}(0)  &  =\frac{1}{2n^{2}N}\sum_{q^{\prime},k,m,k^{\prime
},m^{\prime}}\mu_{m}\mu_{m^{\prime}}\left(  P_{q^{\prime},k,m,q^{\prime
}-q,k^{\prime},m^{\prime}}-Q_{q^{\prime},k,m,q^{\prime}-q,k^{\prime}%
,m^{\prime}}\right)  \times\nonumber\\
&  \times\left(  \frac{1}{E_{q^{\prime},k}+E_{q^{\prime}-q,k^{\prime}}%
}\right)  . \label{susstaticT=0}%
\end{align}

\begin{figure}
[ptb]
\begin{center}
\includegraphics[width=0.5\columnwidth
]%
{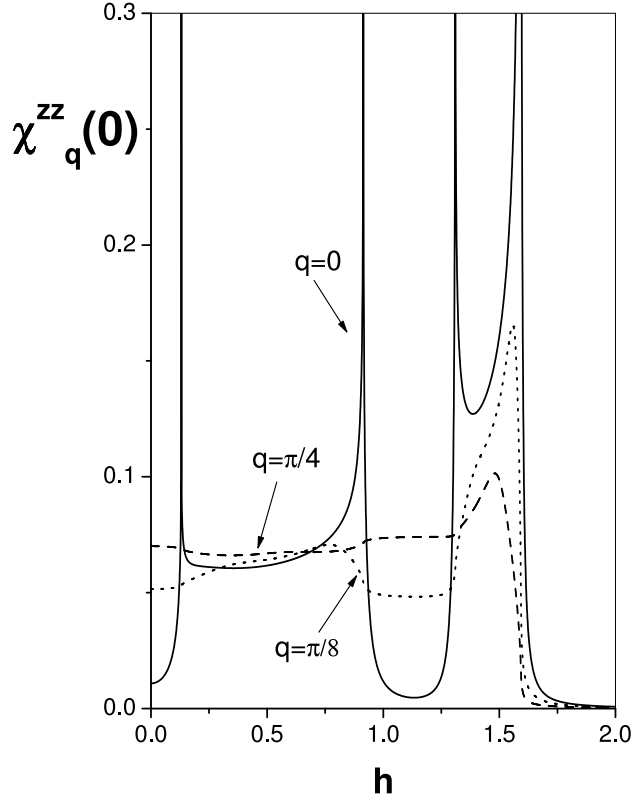}%
\caption{Static susceptibility in the field direction, $\chi_{q}^{zz}%
(0),$ at $T=0,$ as a function of the field for $n=4$, $\mu_{1}=\mu
_{2}=\mu_{3}=\mu_{4}=1,$ $J_{1}=1,$ $J_{2}=1.5,$ $J_{3}=1.75,$ $J_{4}%
=2\ ,\gamma=0.1$\ and different values of $q.$ }%
\label{fig12}%
\end{center}
\end{figure}

The static susceptibility $\chi_{q}^{zz}(0),$ at $T=0$ and for $n=4,$ as a
function of the field is presented in Fig.(\ref{fig12}). For $q=0,$ as
expected, it diverges at the critical fields and it tends to zero as
$h\rightarrow\infty$, which corresponds to the saturation of the model.
However, for $q\neq0$ differently from the isotropic model, no divergence is
present, even for a wave-vector at the zone boundary.

\begin{figure}
[ptb]
\begin{center}
\includegraphics[
width=0.5\columnwidth
]%
{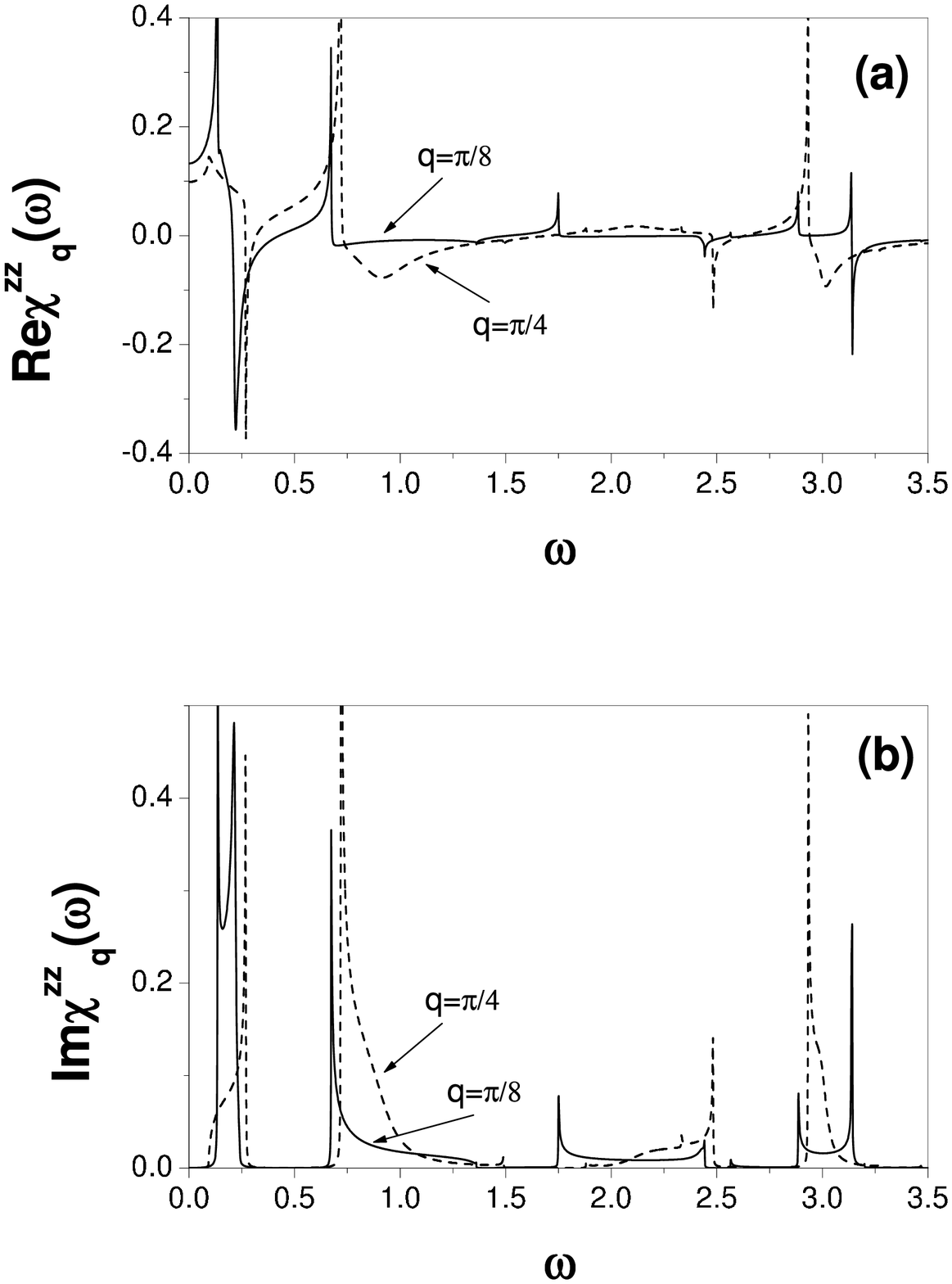}%
\caption{(a) The real and (b) imaginary parts of the dynamic susceptibility
$\chi_{q}^{zz}(\omega),$ at $T=0,$ for $n=4,$ $\mu_{1}=\mu_{2}=\mu_{3}=\mu
_{4}=1,$ $J_{1}=1,$ $J_{2}=1.5,$ $J_{3}=1.75,$ $J_{4}=2,h=0.5$ and
$\gamma=0.1,$ as function of frequency $\omega$ and for different values of
$q.$}%
\label{fig13}%
\end{center}
\end{figure}

The real and imaginary parts of $\chi_{q}^{zz}(\omega)$ are obtained by
considering $\chi_{q}^{zz}(\omega-i\epsilon)$ in the limit $\epsilon
\rightarrow0$ in $\ $eqs.(\ref{susceptibilidade_dinamica-aniso}) and
(\ref{susceptibilidade_estatica-aniso-kubo}). The results, at $T=0$ and for
$n=4,$ are shown in Fig. (\ref{fig13})\textbf{ }for different wave-vectors, as
functions of $\omega.$ As expected, since the Hamiltonian of the model preserves the symmetry of the spin interactions of the  homogeneous one,  the divergences in
the real part, for any wave-vector, correspond to square-root singularities in the imaginary part.

\bigskip%

%

%

\section{Conclusions}

In this work we have considered the anisotropic XY-model on the inhomogeneous
periodic chain with $N$ cells and $n$ sites per cell. The model has been
exactly solved, at arbitrary temperature, for the general case where we have
$n$ different exchange constants and $n$ different magnetic moments. The
number of branches of the excitation spectrum is equal to $n$, and analytical
results can be found for $n\leqq4.$ For $n=2,$ explicit expressions are
presented for the excitation spectrum$,$ and, for $n=8,$ it is obtained numerically.

The induced \ magnetization $M^{z}$ and the isothermal susceptibility
$\chi_{T}^{zz}$ are also given by explicit expressions at arbitrary
temperature. At $T=0$, where the quantum transitions induced by the tranverse
field occur, the spontaneous magnetization $M^{x}$ is written in terms of the
asymptotic behaviour of the determinant of a Toeplitz matrix which corresponds
to the static two-spin correlation in the $x$ direction, at large separation.
The critical behaviour is determined, in high numerical accuracy, by adjusting
the numerical results obtained for finite Toeplitz matrices by using a
non-linear regression of the scaling relation and the $\varepsilon-$algorithm.
This is a rather remarkable result since it allows us to determine very
precisely the critical behaviour of the system within a numerical approach. As
expected, we have shown from these results that the inhomogeneous model
belongs to the same universality class of the homogeneous one.

For $n$ greater than one, the system can present multiple phase transitions
and it is shown that the divergence of the isothermal susceptibility $\chi
_{T}^{zz},$ which is associated to inflection points in the induced
magnetization, is also a signature of the quantum transitions. These critical
points, as usual, correspond to the the points where the spontaneous
magnetization, which is the order parameter characterizing the quantum
transition, goes to zero.

Explicit results are presented for $n=4$ and $n=5,$ where we show that,
differently from the isotropic model where we always have $n$ transitions, the
number of quantum transitions is dependent on the values of the anisotropy
parameters $\gamma.$ We have also concluded that, for $n$ even, there are two
critical $\gamma^{\prime}s$, whereas, for $n$ odd, there just one critical
$\gamma$ equal to zero, and this constitutes the\ main effect of the cell size
on the critical quantum behaviour of the anisotropic model.

It has also been shown that the static two-spin correlation in the $z$
direction, as in the homogeneous model, can present oscillatory or monotonic
behaviour depending on the values of $\gamma^{\prime}s$ and the field. The
limiting surfaces, in the parameter space, which separates the two regimes,
are the so-called disorder surfaces. For the special case when we have
different $J^{\prime}s$ and identical $\gamma$'$s,$ the disorder surfaces
collapse into disorder lines and, in this case, they have been determine
analytically. In particular, for $n=2$, it has been proved exactly that these
lines corresponds to regions where the model can be mapped onto an equivalent
isotropic model. It should be noted that along the disorder lines the direct
static correlation $\left\langle \tau_{l}^{z}(0)\tau_{l+r}^{z}(0)\right\rangle
-\left\langle \tau_{l}^{z}\right\rangle ^{2}$ presents an exponential
asymptotic behaviour, differently from the homogeneous model where it is equal
to zero.

We have also obtained the static and dynamic correlation on the field
directions and, from these results, we have determined the dynamic wave-vector
dependent susceptibility $\chi_{q}^{zz}(\omega).$ Finally, we have shown that,
as in the homogeneous model, the static suceptibility $\chi_{0}^{zz}(0)$ is
equal to the isothermal one $\chi_{T}^{zz}.$

\section{Acknowledgements}

The authors would like thank the Brazilian agencies CNPq, Capes and Finep for
partial financial support, and Dr. A. P. Vieira for enlightening discussions.
L. L. Gon\c{c}alves would like to thank \ Prof. S. R. A. Salinas for his
hospitality during his stay, on a sabbatical leave, at the Departamento de
F\'{\i}sica Geral, Instituto de F\'{\i}sica, Universidade de S\~{a}o Paulo,
when this work was initiated.

\section{Appendix}

In this appendix we present analitycally the mapping of the anisotropic model
onto the isotropic one, for a given set of parameters, for a chain with $n=2$.
This mapping will be obtained, as in the homogeneous model \cite{barouch:1971}%
, by imposing that the isotropic and the anisotropic models have the same
spectrum. As it will be shown, this equivalence is only possible when
$\gamma_{1}=\gamma_{2}.$

Therefore, let us consider the matrices $\mathbb{A}_{q}$ and$\mathbb{\ B}%
_{q},$ given by eqs.(\ref{Aq}) and (\ref{Bq}), which in this case are
explicitly given by
\begin{equation}
\mathbb{A}_{q}=-\left(
\begin{array}
[c]{cc}%
\mu_{1}h & \frac{J_{1}}{2}+\frac{J_{2}}{2}\exp(-i2q)\\
\frac{J_{1}}{2}+\frac{J_{2}}{2}\exp(i2q) & \mu_{2}h
\end{array}
\right)  ,
\end{equation}%
\begin{equation}
\mathbb{B}_{q}=-\left(
\begin{array}
[c]{cc}%
0 & \frac{J_{1}\gamma_{1}}{2}-\frac{J_{2}\gamma_{2}}{2}\exp(-i2q)\\
-\frac{J_{1}\gamma_{1}}{2}+\frac{J_{2}\gamma_{2}}{2}\exp(i2q) & 0
\end{array}
\right)  .
\end{equation}
From these results we obtain the matrix $\left(  \mathbb{A}_{q}%
+\mathbb{B}_{q}\right)  \left(  \mathbb{A}_{q}-\mathbb{B}_{q}\right)  ,$ whose eigenvalues, determined from eq.(\ref{spectrum}), which constitute the two
branches of the spectrum of the chain which are given by%
\begin{align}
E_{q,k=1,2}^{2}  &  =\frac{1}{2}\left[  \alpha_{1}+\alpha_{2}+\left(
c_{1}+c_{2}\right)  \cos2q\right]  \pm\frac{1}{2}\left\{  \left[  \alpha
_{1}+\alpha_{2}+\left(  c_{1}+c_{2}\right)  \cos2q\right]  ^{2}\right.
-\nonumber\\
&  -\left.  4\left[  \left(  \alpha_{1}+c_{1}\cos2q\right)  \left(  \alpha
_{2}+c_{2}\cos2q\right)  -\beta^{2}-\delta^{2}-2\beta\delta\cos2q\right]
\right\}  ^{1/2}, \label{spec2aniso}%
\end{align}
where%
\begin{align}
\alpha_{1}  &  \equiv\mu_{1}^{2}h^{2}+a_{1}^{2}+b_{2}^{2};\text{ }\alpha
_{2}\equiv\mu_{2}^{2}h^{2}+a_{2}^{2}+b_{1}^{2},\nonumber\\
a_{1}  &  \equiv\frac{J_{1}}{2}\left(  1+\gamma_{1}\right)  ;\text{ \ }%
a_{2}\equiv\frac{J_{2}}{2}\left(  1+\gamma_{2}\right)  ,\nonumber\\
b_{1}  &  \equiv\frac{J_{1}}{2}\left(  1-\gamma_{1}\right)  ;\text{ \ }%
b_{2}\equiv\frac{J_{2}}{2}\left(  1-\gamma_{2}\right)  ,\nonumber\\
c_{1}  &  \equiv\frac{J_{1}J_{2}}{2}\left(  1+\gamma_{1}\right)  \left(
1-\gamma_{2}\right)  \text{; \ }c_{2}\equiv\frac{J_{1}J_{2}}{2}\left(
1-\gamma_{1}\right)  \left(  1+\gamma_{2}\right)  ,\nonumber\\
\beta &  \equiv(\mu_{1}b_{1}+\mu_{2}a_{1})h;\text{ \ }\delta\equiv(\mu
_{1}a_{2}+\mu_{2}b_{2})h.
\end{align}
From the previous result we can obtain the spectrum of the isotropic chain,
with parameters $\mu_{1}^{\prime}h^{^{\prime}}$, $\mu_{2}^{\prime}h^{^{\prime
}}$, $J_{1}^{^{\prime}},$ $J_{2}^{^{\prime}},$ by making \ $\gamma_{1}%
=\gamma_{2}=0$ and from this we can write
\begin{align}
\varepsilon_{q,k=1,2}  &  =\frac{1}{2}\left[  (\mu_{1}^{\prime2}+\mu
_{2}^{\prime2})h^{^{\prime}2}+\frac{J_{1}^{^{\prime}2}}{2}+\frac
{J_{2}^{^{\prime}2}}{2}+J_{1}^{^{\prime}}J_{2}^{^{\prime}}\cos2q\right]
\pm\nonumber\\
&  \pm\frac{1}{2}\left(  \mu_{1}^{\prime}+\mu_{2}^{\prime}\right)
h^{^{\prime}}\sqrt{\left(  \mu_{1}^{\prime}-\mu_{2}^{\prime}\right)
^{2}h^{^{\prime}2}+J_{1}^{^{\prime}2}+J_{2}^{^{\prime}2}+2J_{1}^{^{\prime}%
}J_{2}^{^{\prime}}\cos2q}. \label{spec2iso}%
\end{align}
By imposing the equivalence of the two spectra we conclude immediately that we
should have $c_{1}=c_{2}\equiv c$, which implies that $\gamma_{1}=\gamma
_{2}\equiv\gamma.$ Introducing these conditions in eq.(\ref{spec2aniso}) we
obtain
\begin{equation}
E_{q,k=1,2}^{2}=\frac{1}{2}\left[  \alpha_{1}+\alpha_{2}+2c\cos2q\right]
\pm\frac{1}{2}\sqrt{\left(  \alpha_{1}-\alpha_{2}\right)  ^{2}+4\left(
\beta^{2}+\delta^{2}\right)  +8\beta\delta\cos2q,} \label{specanisogamaigual}%
\end{equation}
and by comparing it with eq.(\ref{spec2iso}) we can write immediately the
mapping relations%
\begin{gather}
(\mu_{1}^{\prime2}+\mu_{2}^{\prime2})h^{^{\prime}2}+\frac{J_{1}^{^{\prime}2}%
}{2}+\frac{J_{2}^{^{\prime}2}}{2}=(\mu_{1}^{2}+\mu_{2}^{2})h^{2}+\frac{\left(
J_{1}^{2}+J_{2}^{2}\right)  }{2}\left(  1+\gamma^{2}\right)  ,\label{1}\\
J_{1}^{^{\prime}}J_{2}^{^{\prime}}=J_{1}J_{2}\left(  1-\gamma^{2}\right)
,\label{2}\\
\left(  \mu_{1}^{\prime2}-\mu_{2}^{\prime2}\right)  ^{2}h^{^{\prime}%
4}+h^{^{\prime}2}\left(  \mu_{1}^{\prime}+\mu_{2}^{\prime}\right)  ^{2}\left(
J_{1}^{^{\prime}2}+J_{2}^{^{\prime}2}\right)  =\left(  \mu_{1}^{2}-\mu_{2}%
^{2}\right)  ^{2}h^{4}+\nonumber\\
+\gamma^{2}\left(  J_{1}^{2}-J_{2}^{2}\right)  ^{2}+\left(  J_{1}^{2}%
+J_{2}^{2}\right)  \left[  \left(  \mu_{1}+\mu_{2}\right)  ^{2}h^{2}%
-\gamma^{2}\left(  \mu_{1}-\mu_{2}\right)  ^{2}h^{2}\right]  ,\\
\left(  \mu_{1}^{\prime}+\mu_{2}^{\prime}\right)  ^{2}h^{^{\prime}2}%
J_{1}^{^{\prime}}J_{2}^{^{\prime}}=J_{1}J_{2}\left[  \left(  \mu_{1}+\mu
_{2}\right)  ^{2}h^{2}-\gamma^{2}\left(  \mu_{1}-\mu_{2}\right)  ^{2}%
h^{2}\right]  . \label{4}%
\end{gather}
This set of equations is not independent, and this means that the parameters
of the equivalent isotropic model are not uniquely determined. Moreover, we
can show that eqs.(\ref{1}) to (\ref{4}) can only present a solution if the
parameters of the anisotropic model satisfy the equation%

\begin{equation}
\left(  J_{1}^{2}-J_{2}^{2}\right)  ^{2}\left(  1-\gamma^{2}\right)  ^{2}%
-8\mu_{1}\mu_{2}h^{2}\left(  J_{1}^{2}+J_{2}^{2}\right)  \left(  1-\gamma
^{2}\right)  +16\mu_{1}^{2}\mu_{2}^{2}h^{4}=0. \label{4a}%
\end{equation}
This expression is obtained after some lengthy but straigthforward algebra, by
using the result
\begin{equation}
\left(  \mu_{1}^{\prime}+\mu_{2}^{\prime}\right)  ^{2}h^{^{\prime}2}%
=\frac{\left(  \mu_{1}+\mu_{2}\right)  ^{2}h^{2}-\gamma^{2}\left(  \mu_{1}%
-\mu_{2}\right)  ^{2}h^{2}}{1-\gamma^{2}}, \label{5}%
\end{equation}
which is obtained from eqs. (\ref{1}) and (\ref{4}), and eqs.\ (\ref{1}%
),(\ref{2}) e (\ref{5}). The solution of eq.(\ref{4a}) gives the result%
\begin{equation}
h=\pm\frac{\left\vert J_{1}\pm J_{2}\right\vert \sqrt{1-\gamma^{2}}}%
{2\sqrt{\mu_{1}\mu_{2}}}\equiv\widetilde{h}_{c}\sqrt{1-\gamma^{2}}\label{fin}%
\end{equation}
where $\widetilde{h}_{c}$ are the critical fields of the anisotropic model in
the limit $\gamma\longrightarrow0.$ Although this result, for the disorder lines, is restricted to
$n=2$, we have verified numerically that, for identical $\gamma^{\prime}s,$ it
is still valid for arbitrary $n$. In particular, for n=4, the numerical verification of  eq.(\ref{fin}) is shown in Fig.(\ref{fig8}).

\end{document}